\documentclass{IEEEtaes}

\usepackage[colorlinks,urlcolor=blue,linkcolor=blue,citecolor=blue]{hyperref}

\usepackage{color,array,amsthm}

\usepackage{graphicx}

\usepackage{cite}
\usepackage{amsmath,amssymb,amsfonts}
\usepackage{algorithmic}
\usepackage{textcomp}
\usepackage{verbatim}
\usepackage{xcolor}
\usepackage{units}
\usepackage{subfigure}
\usepackage{comment}
\usepackage{bm}
\usepackage{multirow}

\DeclareMathOperator{\Tr}{Tr}
\def\BibTeX{{\rm B\kern-.05em{\sc i\kern-.025em b}\kern-.08em
    T\kern-.1667em\lower.7ex\hbox{E}\kern-.125emX}}

\newcommand{\inlinerot}[2]{\tiny$\raisebox{1.8pt}{$\underset{ #1 \rightarrow #2 }{}$}$  \hspace{-.87cm}\small $\raisebox{2pt}{${\mathbf{R}}$}$ \hspace{.47cm} \normalsize \hspace{-.11cm}}

\newcommand{\rot}[2]{\underset{#1\rightarrow #2}{{\mathbf{R}}}}

\newcommand{\vecstyle}[1]{{\bm{#1}}}
\newcommand{\matstyle}[1]{{\mathbf{#1}}}
    
\jvol{XX}
\jnum{XX}
\jmonth{XXXXX}
\paper{1234567}
\pubyear{2022}
\doiinfo{TAES.2020.Doi Number}

\begin{document}

\title{Autonomous Asteroid Characterization Through Nanosatellite Swarming}

\author{KAITLIN DENNISON}
\affil{Stanford University, Stanford, CA, 94305, USA} 

\author{NATHAN STACEY}
\affil{Stanford University, Stanford, CA, 94305, USA}

\author{SIMONE D'AMICO}
\member{Member, IEEE}
\affil{Stanford University, Stanford, CA, 94305, USA}


\receiveddate{This work is supported by the NASA Small Spacecraft Technology Program cooperative agreement number 80NSSC18M0058 for contributions to the Autonomous Nanosatellite Swarming (ANS) Using Radio-Frequency and Optical Navigation project. This work is partially supported by the Air Force Office of Scientific Research (AFOSR) via Centauri under the project titled Modular State-Adaptive Landmark Tracking. Funding was received through Air Force Research Laboratory (AFRL) contract \#FA9451-16-D-0014. Fellowship financial support was also provided by the Achievement Rewards for College Scientists (ARCS) Foundation.} 

\corresp{\itshape (Corresponding author: K. Dennison)}

\authoraddress{Kaitlin Dennison, Nathan Stacey, and Simone D'Amico are with the Dept. of Aeronautics \& Astronautics at Stanford University, Stanford, CA 94305 (e-mails: kdenn@stanford.edu, nstacey@stanford.edu, and damicos@stanford.edu).} 


\markboth{DENNISON ET AL.}{AUTONOMOUS ASTEROID CHARACTERIZATION THROUGH NANOSATELLITE SWARMING}
\maketitle

\begin{abstract}This paper first defines a class of estimation problem called simultaneous navigation and characterization (SNAC), which is a superset of simultaneous localization and mapping (SLAM). A SNAC framework is then developed for the Autonomous Nanosatellite Swarming (ANS) mission concept to autonomously navigate about and characterize an asteroid including the asteroid gravity field, rotational motion, and 3D shape. The ANS SNAC framework consists of three modules: 1) multi-agent optical landmark tracking and 3D point reconstruction using stereovision, 2) state estimation through a computationally efficient and robust unscented Kalman filter, and 3) reconstruction of an asteroid spherical harmonic shape model by leveraging a priori knowledge of the shape properties of celestial bodies. Despite significant interest in asteroids, there are several limitations to current asteroid rendezvous mission concepts. First, completed missions heavily rely on human oversight and Earth-based resources. Second, proposed solutions to increase autonomy make oversimplifying assumptions about state knowledge and information processing. Third, asteroid mission concepts often opt for high size, weight, power, and cost (SWaP-C) avionics for environmental measurements. Finally, such missions often utilize a single spacecraft, neglecting the benefits of distributed space systems. In contrast, ANS is composed of multiple autonomous nanosatellites equipped with low SWaP-C avionics. The ANS SNAC framework is validated through a numerical simulation of three spacecraft orbiting asteroid 433 Eros. The simulation results demonstrate that the proposed architecture provides autonomous and accurate SNAC in a safe manner without an a priori shape model and using only low SWaP-C avionics.
\end{abstract}


\section{INTRODUCTION} 
Asteroids and other small bodies provide insight into the formation of the solar system and may hold clues to the origin of life\cite{council_vision_2011, anders_pre-biotic_1989}. Further study of asteroids may enable utilization of their resources and prevention of future asteroid-Earth collisions\cite{coradini_vesta_2011, adams_double_2019}. Reducing mission cost and Earth-based resource dependence is crucial to enable a greater number of future asteroid rendezvous missions. 

To address these needs, this paper defines a new class of estimation problem called simultaneous navigation and characterization (SNAC). In SNAC, the navigation of an agent is coupled with the estimation of properties of the environment or other agents. Simultaneous localization and mapping (SLAM) is similar, but the estimated properties are restricted to a physical map of the environment \cite{thrun_probabilistic_2005}. Thus, SNAC is a superset of SLAM. In the context of this work specifically, SNAC includes estimating the asteroid gravity field, global shape, and rotational motion in addition to the spacecraft states. These SNAC estimates enable safe and fuel efficient operations in proximity to the asteroid, facilitate landings, and yield important scientific insights. This paper develops the SNAC algorithmic framework of the Autonomous Nanosatellite Swarming (ANS) mission concept to autonomously characterize an asteroid using multiple small spacecraft\cite{stacey_autonomous_2018, dennison_comparing_2021}.




Asteroid missions to date such as NEAR Shoemaker\cite{williams_technical_2002}, OSIRIS-REx\cite{williams2018osiris}, and the Hayabusa missions \cite{hashimoto_vision-based_2010, watanabe_hayabusa2_2019} have successfully navigated about and characterized asteroids. However, they have depended extensively on human oversight and highly solicited Earth-based resources such as the NASA Deep Space Network (DSN) for frequent radiometric tracking and communication with the spacecraft. This approach results in expensive mission operations and further burdens already highly subscribed Earth-based resources.

Onboard sensor data is typically downlinked to the ground and combined with the radiometric tracking data in order to perform navigation and characterization \cite{williams_technical_2002, watanabe_hayabusa2_2019, williams2018osiris}. Navigation and characterization are typically achieved using offline algorithms that are too computationally expensive for onboard implementation such as batch orbit determination techniques\cite{konopliv_global_2002} and stereo-photoclinometry (SPC)\cite{gaskell_characterizing_2008,mastrodemos_optical_2012}. Batch orbit determination estimates the spacecraft states and can include the estimation of sensor biases as well as properties of the target body such as parameters describing the gravity field and rotational motion \cite{konopliv_global_2002,miller_determination_2002}. SPC generates a highly-detailed digital terrain model (DTM) of the target from images and other sensor data if available \cite{gaskell_characterizing_2008}. The DTM can be used to aid gravity recovery and state estimation as well as landmark tracking for optical navigation \cite{williams_technical_2002,watanabe_hayabusa2_2019,williams2018osiris}. When relying on Earth-based resources and human oversight, the spacecraft is slow to react to its environment due to delays between recording sensor data and receiving commands from the ground. Thus for safety reasons, caution must be exercised when reducing the spacecraft altitude, which can limit the time spent in lower-altitude orbits where higher-resolution data is obtained. 

Current proposed solutions to increase autonomy in asteroid rendezvous missions often make oversimplifying assumptions, rely on an a priori DTM, or increase hardware power and cost requirements. Examples of assumptions include idealized image processing, time-keeping, and estimation as well as the absence of process noise and sensor biases \cite{stacey_autonomous_2018,leonard_absolute_2012,hesar_small_2015,bhaskaran_small_2011,fujimoto_stereoscopic_2016,atchison2}. Such assumptions can lead to significant state estimation biases, as occurred with Orbital Express \cite{dennehy2011summary}, or even system failure, as seen in the Demonstration of Autonomous Rendezvous Technology (DART) mission \cite{croomes2006overview}. Furthermore, some autonomous mission concepts like \cite{hesar_small_2015} and \cite{cheng_autonomous_2003} use a DTM generated prior to autonomous navigation but that typically requires a process like SPC to be performed. To avoid an a priori DTM model, LIDAR is often used to measure depth and build a surface point cloud onboard \cite{dietrich2014asteroid,woods2016lidar,williams_technical_2002}. However, LIDAR is a high size, weight, power, and cost (SWaP-C) component. Moreover, scanning LIDAR introduces moving parts, which means more points of failure, and flash LIDAR reduces maximum observation range, which limits the spacecraft's altitude \cite{driedger2022feasibility}.



ANS overcomes these limitations by utilizing an autonomous swarm of small satellites that cooperate to characterize an asteroid using intersatellite radio-frequency (RF) measurements, multi-agent stereovision, and optical landmark tracking \cite{stacey_autonomous_2018}. The ANS SNAC framework consists of three novel modules. One, the landmark tracking and stereovision module performs multi-agent optical landmark tracking and 3D point reconstruction. This enables LIDAR-free depth estimation and optical navigation without an a priori DTM via multi-agent stereovision. Two, the state estimation filter module estimates the spacecraft states, relative clock offsets, and asteroid properties through a computationally efficient and robust unscented Kalman filter (UKF). Three, the global shape reconstruction module generates a spherical harmonic shape model of the asteroid through a new technique that leverages a priori empirical knowledge of the shape properties of celestial bodies. The spherical harmonic shape model can be computed onboard and used in place of a highly-detailed DTM.

The algorithmic pipeline is made possible and enhanced by a distributed space system, which consists of multiple spacecraft cooperating to complete a shared objective. The advantages of this architecture include distributed computation as well as greater robustness, accuracy, flexibility, sensing coverage, and redundancy. For these reasons, distributed space systems are becoming more popular and often leverage advances in small spacecraft technology to reduce cost \cite{klesh_marco_2018, tapley_gravity_2004, sanchez_starling1_2018, koenig_formation_2021}.

The contributions of this paper significantly advance the authors' prior work on ANS \cite{stacey_autonomous_2018,dennison_comparing_2021}. This work is also an update to the conference paper of the same title \cite{stacey_dennison_2022} and includes major improvements to the landmark tracking and stereovision module.

Following this introduction, Section \ref{sec:conops} details the ANS concept of operations. Section \ref{sec:ldt} describes the process for optically tracking landmarks and reconstructing them in 3D space using multi-agent stereovision. Section \ref {sec:estimation} describes how optical pixel measurements of landmarks are fused with intersatellite RF measurements in a computationally efficient and robust UKF to estimate the spacecraft states, relative clock offsets, and asteroid properties. Section \ref{sec:shape} presents the novel global asteroid shape reconstruction technique. Section \ref{sec:validation} validates the ANS SNAC algorithmic architecture through the numerical simulation of three spacecraft orbiting the asteroid 433 Eros. Finally, Section \ref{sec:conclusion} draws conclusions based on the numerical results.



\section{CONCEPT OF OPERATIONS}\label{sec:conops} 
\begin{figure}[!ht]
\centering
\includegraphics[width=8cm,trim=0 0 0 0,clip]{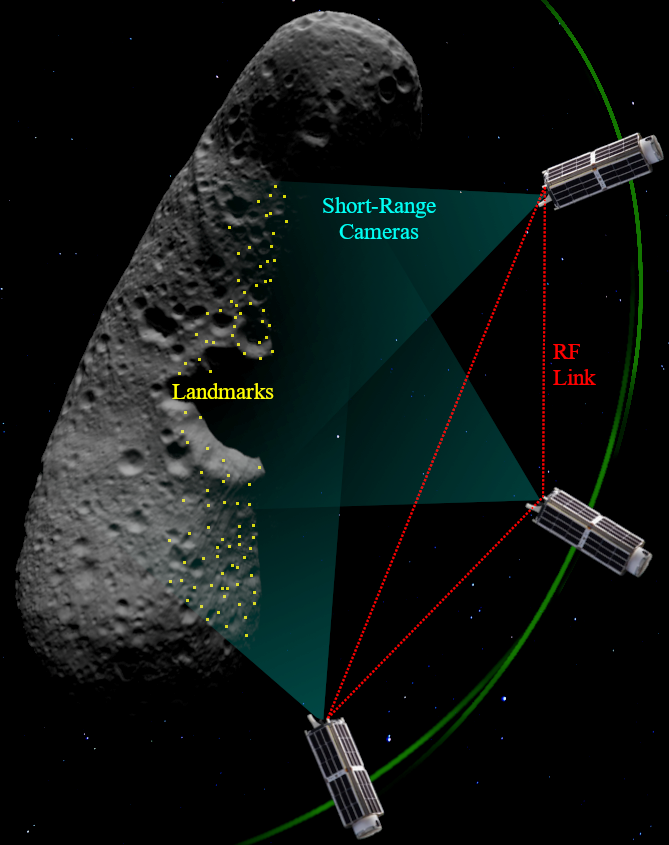}
\caption{Conceptual illustration of ANS including the spacecraft orbits, intersatellite RF links, and landmarks tracked using the spacecraft's short-range cameras.}
\label{fig:ConOps}
\end{figure}

The ANS mission concept consists of a mothership and one or more deputy nanosatellites, all of which cooperate to characterize an asteroid via SNAC using only low SWaP-C hardware (see Fig. \ref{fig:ConOps}). The spacecraft are in closed, reconfigurable orbits about the target asteroid, which provides a geometrically diverse set of measurements over a long period of time. 

\begin{figure}[!ht]
\centering
\includegraphics[width=8cm,trim=0 0 0 0,clip]{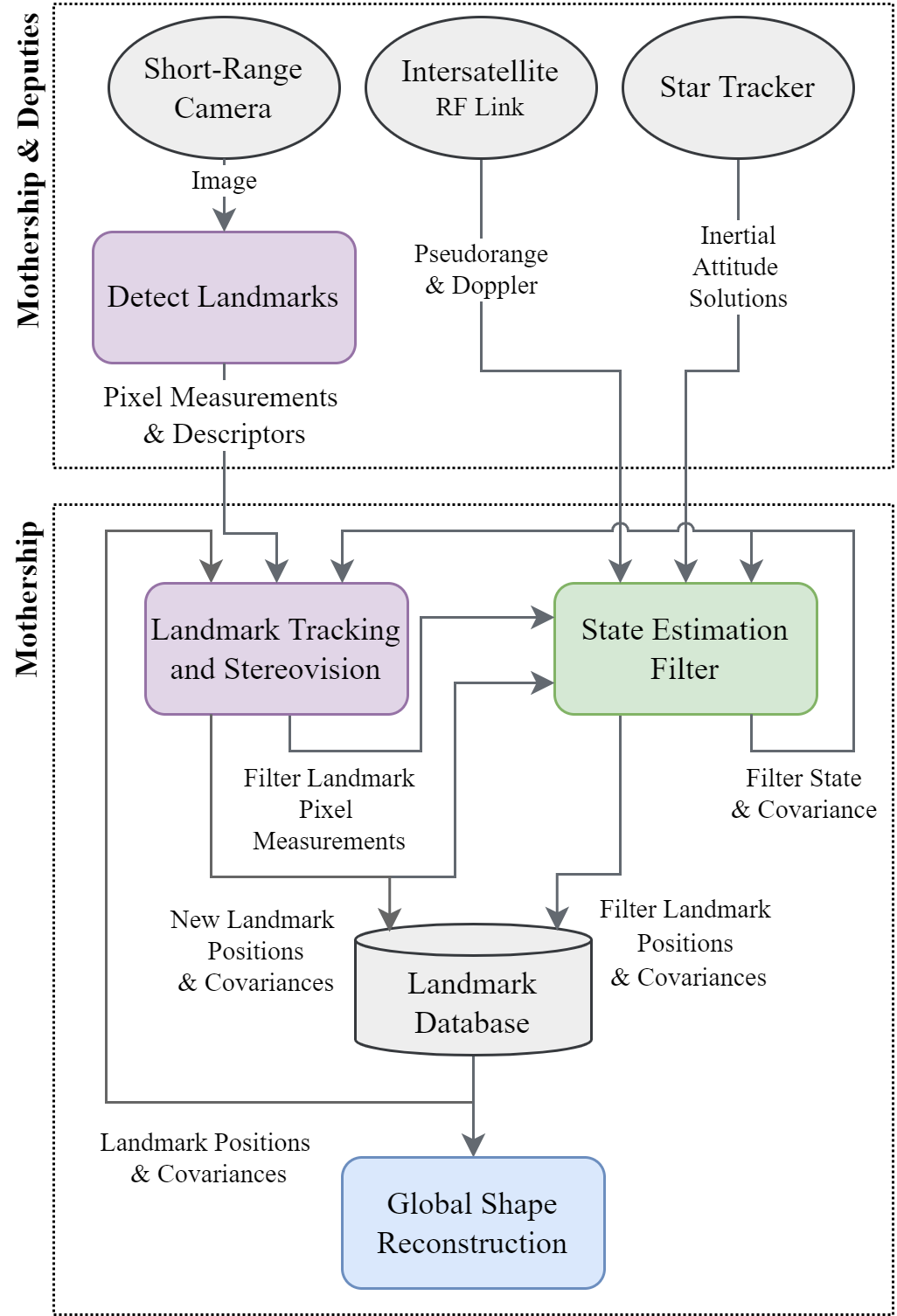}
\caption{ANS data transfer and system interactions to perform SNAC. The ovals are sensors, the cylinder is data storage, and the rectangles are algorithms. Each algorithmic module is a different color.}
\label{fig:flowchart}
\end{figure}

The mothership carries the deputies to the target asteroid and is a larger spacecraft. The mothership has greater computational resources and is the only spacecraft equipped with a high-gain antenna to communicate with the Earth, primarily at the beginning and end of the mission and at major intermediate milestones. Although some computational tasks are executed exclusively on the mothership, the image processing is distributed among the swarm in order to reduce the maximum computational burden on any one spacecraft. The proposed algorithmic modules and intersatellite data transfer necessary to perform SNAC within the ANS architecture are depicted in Fig. \ref{fig:flowchart}. 

The low SWaP-C sensors onboard each spacecraft include a short-range camera, a star-tracker, a chip scale atomic clock, and an intersatellite RF link. Potential small spacecraft avionics can be found in \cite{yost_smallspacecraft_2021} and are discussed further in Section \ref{sec:validation}. The short-range cameras are used to optically detect and track asteroid landmarks. The star trackers provide inertial attitude solutions for each spacecraft. The intersatellite RF links between every pair of spacecraft provide intersatellite one-way pseudorange and Doppler measurements while facilitating communication and data transfer. 

Data transfer from each deputy spacecraft to the mothership includes the intersatellite pseudorange and Doppler measurements, inertial attitude solutions, and landmark pixel measurements and feature descriptors. Landmark detection and inertial attitude solutions are computed by each individual spacecraft, distributing computational load amongst the swarm. The primary data transferred from the mothership to the deputies relates to maneuver commands. However, guidance and control are not considered in this paper, and therefore are not illustrated in Fig. \ref{fig:flowchart}. For more information on guidance and control for ANS, see \cite{lippe_spacecraft_2020}. 

The autonomously-populated landmark database is stored on the mothership and is depicted by the cylinder in Fig. \ref{fig:flowchart}. The landmark database includes the 3D asteroid-centered asteroid-fixed (ACAF) positions and associated error covariances of landmarks that are currently tracked in the filter state as well as landmarks that have been retired (i.e., removed and stored) from the filter state. The coordinate frames used throughout this paper are defined in Table \ref{tab:frames}.

\begin{table}
\caption{Coordinate frame definitions. If unspecified, the third axis completes the right-handed triad.}
\label{tab:frames}
\centering
\setlength{\tabcolsep}{2.5pt}
\begin{tabular}{ll}
\hline
\bfseries Frame & \bfseries Definition\\
\hline
ACAF            &Asteroid-Centered Asteroid-Fixed\\
                &X-axis aligned with asteroid prime meridian\\
                &Z-axis aligned with asteroid mean spin axis\\
\hline
ACI             &Asteroid-Centered Inertial \\
                &X-axis aligned with vernal equinox\\
                &Z-axis normal to Earth mean-equatorial plane\\
\hline
ACIC            &Asteroid-Centered Inertial Control\\
                &X-axis aligned with ACAF x-axis at J2000 epoch\\
                &Z-axis aligned with asteroid mean spin axis\\
\hline
CF              &Camera Fixed\\
                &X-axis parallel to image plane pixel columns\\
                &Y-axis parallel to image plane pixel rows\\
                &Z-axis aligned with camera boresight\\
\hline
RTN             &Radial Transverse Normal\\
                &R-axis aligned with s/c position vector\\
                &N-axis aligned with s/c angular momentum vector\\
\hline
\end{tabular}
\end{table}

The swarm orbit geometry of ANS is designed to provide safe and accurate asteroid characterization. The mean semi-major axis of each spacecraft is equal to prevent the swarm from quickly drifting apart in the along-track direction. The spacecraft are primarily separated in the along-track direction, which balances the competing angular separation requirements of landmark tracking and stereovision; see Section \ref{sec:ldt} for more information. A conceptual depiction of this formation and the spacecraft sensors is shown in Fig. \ref{fig:ConOps}. 


\subsection{SNAC Algorithmic Architecture}

Upon initial approach, a brief ground-in-the-loop phase uses optical navigation and ground-based radiometric tracking to get coarse initial estimates of the spacecraft states, asteroid rotational parameters, and asteroid gravity field spherical harmonic coefficients. This phase can be accomplished similarly to completed missions\cite{mastrodemos_optical_2012,lauretta_osirisrex_2017} and is assumed to have already been done in this paper. After the ground-in-the-loop initialization, the SNAC pipeline begins with an empty landmark database. The swarm autonomously initializes landmark position estimates, tracks landmarks, updates the landmark database, and simultaneously estimates the spacecraft states, asteroid characteristics, and landmark positions. This process is divided into the three SNAC modules as illustrated in Fig. \ref{fig:flowchart}. Each module has a different color.



The first module, shown in purple in Fig. \ref{fig:flowchart}, receives input from the cameras, state estimation filter, and landmark database. It uses this information to detect landmarks, correlate them between spacecraft, and initialize their 3D positions via stereovision. This module also tracks landmarks over time and maintains the landmark database. In turn, landmark initial 3D positions and pixel measurements of already-initialized landmarks are provided to the filter. This module is further detailed in Section \ref{sec:ldt}. 

The second module is the state estimation filter shown in green in Fig. \ref{fig:flowchart}. The inputs to this module are the one-way intersatellite RF measurements obtained through the intersatellite RF links, the spacecraft inertial attitudes provided by the star trackers, and the output from the landmark tracking and stereovision module. Specifically, the required output from the landmark tracking and stereovision module is pixel measurements of landmarks currently in the filter state and initial position estimates of new landmarks to be incorporated into the filter state. The inputs of the state estimation filter module are fused in a UKF that simultaneously estimates the spacecraft states, relative clock offsets, landmark ACAF positions, asteroid rotational motion, and asteroid gravity field coefficients. The state estimates and associated error covariances are provided to the landmark tracking and stereovision module as well as the landmark database. This module is only executed on the mothership and is described in Section \ref{sec:estimation}.

The third module is global shape reconstruction, which is shown in blue in Fig. \ref{fig:flowchart}. The only input is the ACAF landmark positions and associated error covariances stored in the landmark database. This module generates a global asteroid spherical harmonic shape model that can be computed at any time in the mission. This module is only executed on the mothership and is described in Section \ref{sec:shape}.

\section{LANDMARK TRACKING AND STEREOVISION}\label{sec:ldt} 

The landmark tracking and stereovision module handles landmark position initialization, tracking, and retirement to support state estimation and shape generation within SNAC. This module is represented by the purple blocks in Fig. \ref{fig:flowchart}. After landmarks are initialized, state estimation requires the pixel measurements of as many landmarks in the filter state as possible to perform an update. However, as the asteroid rotates and the spacecraft continue in their orbits, landmarks enter and leave the field of view. Thus, new landmarks must be added to the filter and obsolete ones removed. 


This entire process is accomplished per the following five steps.
\begin{enumerate}
    \item Landmarks are detected in the images taken by each spacecraft at the same nominal epoch. \
    
    \item Landmarks are matched between images by spacecraft-to-spacecraft correlation. \ 
    
    \item New landmarks matched through spacecraft-to-spacecraft correlation are initialized in 3D space using multi-agent stereovision. This provides initial position estimates of new landmarks for the filter. \
    
    \item Landmarks currently in the filter are matched to landmarks in the images in filter-to-spacecraft correlation. This serves the need of measurement updates of the UKF. \
    
    \item Landmarks are retired from the filter once they have not been tracked for a set number of consecutive time steps. This prevents excessive memory usage. \
\end{enumerate}
Technically, steps two and three can occur concurrently with steps four and five. Landmark tracking and stereovision uses the filter mean state estimate and associated covariance for the spacecraft positions, asteroid rotation parameters, and landmark positions (post-initialization). The filter output is transferred after the filter time update but before the measurement update so the filter mean state estimate and associated covariance are at the same nominal epoch as the images. Some mathematical preliminaries are introduced in the following subsection before each of the five steps is described in detail.

\subsection{Mathematical Preliminaries}

Two mathematical concepts, the camera projection matrix and Mahalanobis distance, are used throughout multiple of the five steps of landmark tracking and stereovision. They are introduced generally here and their exact use cases are explained in the relevant portions of the rest of the section.

\subsubsection{Camera Projection Matrix} The camera projection matrix transforms a point from an arbitrary, 3D world (W) frame to the 2D image frame of a specific camera \cite[p.~155]{Hartley2003}. For reference, the ACAF, ACI, and CF frames can each take the place of W in the following equations. 

A landmark 3D position $\vecstyle{L}$ is projected from the W frame into the image frame of the $j$th spacecraft using the finite projective camera model,
\begin{equation}\label{eq:SimpleCameraModel}
\vecstyle{l}_{h,j} = \matstyle{M}_j \vecstyle{L}_h.
\end{equation} 
\noindent Here, $\vecstyle{L}_h = [\vecstyle{L}^T,1]^T$. The $h$ subscript denotes the homogeneous form of a point. The image frame projection of $\vecstyle{L}$ in Euclidean and homogeneous coordinates is
\begin{equation}\label{eq:PixelProjection}
\vecstyle{l}_j =
\begin{bmatrix}
u\\
v
\end{bmatrix}, \quad  \vecstyle{l}_{h,j} =
\begin{bmatrix}
uw\\
vw\\
w
\end{bmatrix},
\end{equation}
\noindent where $u$ and $v$ are the x- and y-axis pixel coordinates and $w$ is the common factor. 

Furthermore, $\matstyle{M}_j$ is the projective camera matrix of the $j$th spacecraft, which is defined as
\begin{equation}\label{eq:ProjectiveTransform}
\matstyle{M}_j = \hat{\matstyle K}_j \bigg[\rot{W}{CF_j} \ \ \ \ \vecstyle{t} \bigg],
\end{equation} 
\noindent where $\hat{\matstyle K}_j$ is the $j$th spacecraft's camera calibration matrix, as defined in \cite[p.~157]{Hartley2003}. The rotation matrix from W to spacecraft $j$'s CF is denoted as \inlinerot{W}{CF_j}. The vector $\vecstyle{t}$ is the 3D vector from the origin of CF$_j$ to the origin of W, expressed in the CF$_j$ frame. In this paper, the origin of CF$_j$ is defined as the respective spacecraft position, allowing $\vecstyle{t}$ to be written as
\begin{equation}\label{eq: def t camera model}
    \vecstyle{t} = \rot{W}{CF_j}(-\vecstyle{r}_j).
\end{equation}
Here, $\vecstyle{r}_j$ is the position of the spacecraft center of mass in the W frame.



\subsubsection{Mahalanobis Distance} Mahalanobis distance $m$ is a measure of the distance between the distribution of a random vector and a sample $\vecstyle{s}$ of that vector. It is used as a metric for multivariate outlier rejection and correspondence to account for cross-correlation between elements of a random vector and for when the variance of each element may not be the same \cite{daszykowski2007robust}. 

If a distribution is characterized by mean $\vecstyle{\mu}$ and covariance $\matstyle{\Sigma}$, then $m$ of $\vecstyle{s}$ to the distribution is defined as
\begin{equation}\label{eq:mahalanobis}
    m = \sqrt{(\vecstyle{s}-\vecstyle{\mu})^T \matstyle{\Sigma}^{-1}(\vecstyle{s}-\vecstyle{\mu})}.
\end{equation}

Outlier rejection is performed by eliminating any samples that have less than a $1-p_m$ probability of being observed as a sample from the distribution characterized by $\vecstyle{\mu}$ and $\matstyle{\Sigma}$. Mahalanobis distance is mapped to $p_m$ by the cumulative distribution function ($\text{CDF}$) of $m^2$ as $p_m=1-\text{CDF}(m_t^2)$ where $p_m$ is the probability of observing an $m^2 > m_t^2$. Thus, a specific $p_m$ results in a specific Mahalanobis distance threshold $m_t$ for the distribution. Consequently, a sample with an $m$ above $m_t$ will be rejected. 

There are two relevant distributions used in this paper that allow the $p_m$ to $m_t$ relationships to be simplified. The first is for a 1D Gaussian normal distribution
\begin{equation}\label{eq:mahalanobis_thresh_1D}
    m_{t,1D} = z_{\nicefrac{p_m}{2}},
\end{equation}
\noindent where $z_{\nicefrac{p_m}{2}}$ denotes the z-score such that the $\text{CDF}$ of the standard normal distribution equals $1-\nicefrac{p_m}{2}$ as explained in \cite[p.~326]{navidi2010principles}. The second is a 2D Gaussian normal distribution
\begin{equation}\label{eq:mahalanobis_thresh_2D}
    m_{t,2D} = \sqrt{-2\ln (p_m)}.
\end{equation} 
Eqs. \eqref{eq:mahalanobis_thresh_1D} and \eqref{eq:mahalanobis_thresh_2D} are derived from a chi-squared distribution of $m^2$ with one degree and two degrees of freedom, respectively.

\subsection{Step One: Landmark Detection}

Keypoint detection identifies unique landmarks in each image via scale-invariant feature transform (SIFT) \cite{lowe_distinctive_2004}. Previous studies have shown that SIFT is the most reliable detection method when compared to other keypoint descriptors, including crater tracking, in the asteroid environment where lighting and perspective changes between images can be significant \cite{dennison_comparing_2021,takeishi_evaluation_2015}. 

The keypoints detected along the edge of the illuminated asteroid surface in the image were removed because they undergo large perspective and lighting changes. This is accomplished by thresholding the image to create a binary mask of the illuminated regions. The dark portion of the mask is dilated with a 6 pixel diameter disk, shrinking the illuminated region along the outer edge.

\subsection{Step Two: Spacecraft-to-Spacecraft Correlation}\label{ssec:sc2sc}

Spacecraft-to-spacecraft correlation matches keypoints between images taken by all spacecraft at the same nominal epoch. For each possible pairing of images among the swarm, the keypoints are correlated using the descriptor-matching algorithm devised by Lowe in \cite{lowe_distinctive_2004}. 

In this work, two methods of rejecting outliers are used on the pairs of SIFT keypoints determined from Lowe's method. The first is maximum likelihood estimator sample consensus (MLESAC); the second is the enforcement of the epipolar constraint. Both methods make use of a computer vision concept called epipolar geometry, which relates 2D points from two camera perspectives (see chapter nine of \cite{Hartley2003}).

MLESAC estimates the epipolar geometry directly from the set of pre-matched points and does not rely on knowledge of the camera positions. MLESAC performance depends on measurement noise and Lowe's correlation. This paper uses MATLAB's MLESAC implementation to estimate the essential matrix \cite[p.~257]{Hartley2003} from the set of corresponding points \cite{torr_mlesac_2000}.

The epipolar constraint uses the filter state estimates to compute the epipolar geometry and identify pre-matched pairs that do not fit the geometry. Thus, MLESAC and the epipolar constraint complement each other to improve spacecraft-to-spacecraft correlation when the filter state estimate uncertainties are large and small, respectively. Reliance on state estimates, however, requires special care, as is explained next. 

The epipolar constraint is computed as described in \cite[p.~225]{klancar2017wheeled}: a matched pair of points $\vecstyle{l}_{1}$ and $\vecstyle{l}_{2}$ is considered an outlier when the epipolar constraint is violated. This is determined by the distance $d$ of $\vecstyle{l}_{2}$ from the epipolar line $\vecstyle{\zeta}_2 = [a, b, c]^T$ projected by $\vecstyle{l}_{1}$ into the second image. In this paper, $d$ is defined as
\begin{equation}\label{eq:DistToLine}
    d = \frac{\vecstyle{\zeta}_2^T\vecstyle{l}_{2,h}}{\sqrt{a^2 + b^2}}.
\end{equation}

Typically, pairs are rejected when $d$ is above a tuned, constant threshold. However, the expected error and noise in $d$ are not constant because the camera position, which is used to calculate $\vecstyle{\zeta}_2$, comes from the UKF.

In this paper, rather than have a fixed $d$ threshold for outlier rejection, $d$ is assumed to follow a 1D Gaussian normal distribution and Mahalanobis distance is used to determine outliers. Mahalanobis distance allows the uncertainty of the spacecraft positions to be accounted for as part of the covariance in Eq. \eqref{eq:mahalanobis}. Thus, Eqs. \eqref{eq:mahalanobis} and \eqref{eq:mahalanobis_thresh_1D} are used to eliminate any corresponding pairs where $m > m_{t,1D}$ formulated with $s = d$, $\mu = 0$, and 
\begin{equation}\label{eq:DistCov}
    \Sigma = \frac{\partial d}{\partial \vecstyle{\gamma}} \matstyle{P}_\vecstyle{\gamma} \frac{\partial d}{\partial \vecstyle{\gamma}}^T + \frac{\partial d}{\partial \vecstyle{l_1}} \matstyle{R}_1 \frac{\partial d}{\partial \vecstyle{l_1}}^T + \frac{\partial d}{\partial \vecstyle{l_2}} \matstyle{R}_2 \frac{\partial d}{\partial \vecstyle{l_2}}^T.
\end{equation}
\noindent The filter mean state estimates of the two spacecraft positions are concatenated as $\vecstyle{\gamma} = [\hat{\vecstyle{r}}_1^T \ \hat{\vecstyle{r}}_2^T]^T$. The matrices $\matstyle{P}_\vecstyle{\gamma}$, $\matstyle{R}_1$, and $\matstyle{R}_2$ are the associated covariance matrices of $\vecstyle{\gamma}$, $\vecstyle{l}_1$, and $\vecstyle{l}_2$, respectively. The partial derivatives in Eq. \eqref{eq:DistCov} were computed using MATLAB's symbolic toolbox. 

This paper incorporates a unique spacecraft-to-spacecraft correlation sharing technique that operates in symbiosis with the formation design of ANS. The technique and formation design are both informed by the competing requirements of spacecraft-to-spacecraft correlation and multi-agent stereovision. As intersatellite separation increases, the false negative rate of spacecraft-to-spacecraft correlation increases because the perspective change becomes too large \cite{dennison_comparing_2021}. However, \cite{beder_determining_2006} and \cite{hess-flores_uncertainty_2014} both showed that stereovision uncertainty decreases as intersatellite separation increases. 

To balance these competing interests, the three spacecraft are separated in the along-track direction so that the two outer spacecraft have a wide baseline and the middle spacecraft shares a short baseline with either neighbor. The separation is small enough such that all three spacecraft have overlapping perspectives. With this formation, the spacecraft that are neighbors will have more matches than the two outer spacecraft. If a keypoint from the middle spacecraft was matched to both outer spacecraft, all three keypoints are considered correlated regardless of whether the outer two keypoints correlated directly. This correlation sharing restores the correlations falsely missed between the outer spacecraft. Stereovision then benefits from a greater number of correlations between the widely-separated spacecraft and three pairs of pixel measurements instead of two.

\subsection{Step Three: Landmark Initialization}

Each set of two or more correlated keypoints is reconstructed into the 3D ACAF frame via multi-agent stereovision. The stereovision estimate for each set of corresponding pixel measurements is
\begin{equation}\label{eq:ReprojectionGeneral}
    \tilde{\vecstyle{L}} = \underset{\vecstyle{L}}{\text{argmin}} \sum_{j=1}^N \left\Vert \begin{bmatrix} \matstyle{M}_j^{(1)}\vecstyle{L}_h \\ \matstyle{M}_j^{(2)}\vecstyle{L}_h \end{bmatrix} \frac{1}{\matstyle{M}_j^{(3)}\vecstyle{L}_h}   - \vecstyle{l}_{j} \right\Vert^2 ,
\end{equation}
where $\matstyle{M}_j^{(1)}$ denotes the first row of $\matstyle{M}_j$, and $N$ is the number of spacecraft whose pixel measurements are used for stereovision. In this paper, Eq. \eqref{eq:ReprojectionGeneral} is solved using the Gauss-Newton method where the initial guess is calculated using the homogeneous linear triangulation method from \cite[p.~312]{Hartley2003}. The ACAF landmark position estimates obtained through stereovision are incorporated into the UKF state vector and error covariance as described in Section \ref{sec:estimation}.

\subsection{Step Four: Filter-to-Spacecraft Correlation}\label{ssec:f2sc}

In the fourth step of landmark tracking and stereovision, landmarks from the filter state are correlated to the keypoints in the image captured by each spacecraft. Mahalanobis distance and the squared Euclidean norm of the difference between consecutive keypoint descriptors are used as the correlation metrics. The $i$th filter landmark can have a correlated keypoint in the image from each spacecraft at the current time step but can only correlate to a single keypoint in each image.

The keypoint $k^*$ in the image from the $j$th spacecraft that correlates to filter landmark $i$ is determined through a weighted optimization problem. The optimization problem iterates over all keypoints $k$ in the image and selects the single best $k$ that satisfies the constraints to return as $k^*$. The cost function is a weighted sum of three different Mahalanobis distances and the squared Euclidean norm of the difference between feature descriptor vectors. The constraints enforce outlier rejection. The optimization problem is
\begin{equation}\label{eq:CorrCost}
\begin{array}{lll}
k^* = & \underset{k}{\text{argmin}} & \hat{w}_{\text{2D}}m_{ik} + \hat{w}_{u} m_{ik,u} \vspace{-1.5mm}\\ 
    &                               &  \ + \ \hat{w}_{v} m_{ik,v} \vspace{2mm} + \Vert \vecstyle{f}_i - \vecstyle{f}_k \Vert^2 \\
    & \textrm{subject to}           & m_{ik,\text{2D}} \leq m_{t,\text{2D}} \\
    &                               & m_{ik,u} \leq m_{t,\text{1D}} \\
    &                               & m_{ik,v} \leq m_{t,\text{1D}} \\
    &                               & \Vert \vecstyle{f}_i - \vecstyle{f}_k \Vert^2 \leq \delta_{f,t}.
\end{array}
\end{equation}

Above, $m_{ik}$, $m_{ik,u}$, and $m_{ik,v}$ are, respectively, the 2D Mahalanobis distance, 1D Mahalanobis distance in the image $u$-direction, and the 1D Mahalanobis distance in the image $v$-direction between the filter landmark projected into the image frame and the keypoint. The values $\hat{\omega}_{2D}$, $\hat{\omega}_{u}$, and $\hat{\omega}_{v}$ are user-specified weights. Furthermore, $\vecstyle{f}_i$ and $\vecstyle{f}_k$ are the feature descriptor vectors of the filter landmark and image keypoint, respectively. A user-specified threshold for the descriptor difference is $\delta_{f,t}$. The origin and purpose of each component of Eq. \eqref{eq:CorrCost} is explained in the rest of this subsection.

Mahalanobis distance is used in filter-to-spacecraft correlation for both outlier rejection and correspondence. In Section \ref{ssec:sc2sc}, Mahalanobis distance was used only for outlier rejection and with a completely different formulation of Eq. \eqref{eq:mahalanobis} than any of the three formulations of Mahalanobis distance in Eq. \eqref{eq:CorrCost}.

The 2D Mahalanobis distance formulation $m_{ik}$ is the distance of $\vecstyle{l}_k$ from the expected 2D projection $\hat{\vecstyle{l}}_i$ and associated 2D covariance $\matstyle{P}_{\textit{l,i}}$ of the $i$th landmark's mean state estimate $\hat{\vecstyle{L}}_i$ from the filter. The matrix $\matstyle{P}_{\textit{l,i}}$ is computed through a linear mapping of the error covariances of the observing spacecraft position and the asteroid rotational parameters contained in the filter state. This utilizes the partial derivatives of Eq. \eqref{eq:SimpleCameraModel} with respect to the spacecraft position and asteroid rotational parameters. Accordingly, $m_{ik}$ is computed from Eq. \eqref{eq:mahalanobis} with $\vecstyle{s}=\vecstyle{l}_k$, $\vecstyle{\mu}=\hat{\vecstyle{l}}_i$ and $\matstyle{\Sigma}=\matstyle{P}_{\textit{l,i}}+\matstyle{R}_k$. The value $\matstyle{R}_{k}$ is the measurement error covariance of the $k$th keypoint's pixel center. 

The two other variants of $m_{ik}$ ($m_{ik,u}$ and $m_{ik,v}$) are the 1D Mahalanobis distance of $\vecstyle{l}_k$ from $\hat{\vecstyle{l}}_i$ in the respective $u$- and $v$-directions of the image frame. They are computed using Eq. \eqref{eq:mahalanobis} and the respective scalar component of $\vecstyle{s}$, $\vecstyle{\mu}$, and $\matstyle{\Sigma}$ used for the 2D $m_{ik}$. While $m_{ik}$ is the main source of outlier rejection with $\hat{\omega}_{2D} > \hat{\omega}_{u}$ and $\hat{\omega}_{2D} > \hat{\omega}_{v}$, the 1D versions are included in \eqref{eq:CorrCost} to reject outliers that lie along the $u$- and $v$-axes.

There a caveat when using Mahalanobis distance and keypoint descriptors together: co-location of keypoints can cause miscorrelation. Interesting asteroid surface points tend to cluster (e.g., keypoints detected on a crater rim or a rocky pile). This means another keypoint right next to the true match might be selected simply through Mahalanobis distance because of the noise and uncertainty in state estimation. Feature descriptors provide an additional correlation metric because they are not subject to state estimate uncertainty.

The feature descriptor $\vecstyle{f}_i$ associated with a landmark is retained any time it is correlated in an image. To further eliminate candidate keypoints, the $\vecstyle{f}_i$ from the previous time step is compared to keypoint $k$'s feature descriptor $\vecstyle{f}_k$. The feature descriptors are compared by taking the squared Euclidean norm of their difference. A maximum threshold $\delta_{d,t}$ is imposed on the feature descriptor differences to eliminate outliers. The $\vecstyle{f}_i$ from spacecraft $j$ is used when available but if not, the $\vecstyle{f}_i$ of another spacecraft is used. 

When the $k^*$th keypoint is correlated to the $i$th landmark, the keypoint's centroid $\vecstyle{l}_{k^*}$ is provided to the estimation filter as the pixel measurement associated with the $i$th landmark. This completes the process to correlate keypoint $k^*$ in the image from the $j$th spacecraft to landmark $i$. This process is repeated for all spacecraft and landmarks currently in the filter state. 

\subsection{Step Five: Landmark Retirement}\label{sec:landmark retirement}

The final step of the landmark tracking and stereovison subsystem is to determine which landmarks to retire from the filter state and how to maintain the landmark database. Retired landmarks are retained in the database for use in global shape reconstruction only. They are never used for any correlation method again, nor are their ACAF positions and covariances updated. 

Landmarks are retired because true correlations may not pass the descriptor difference constraint in Eq. \eqref{eq:CorrCost} of landmark tracking if too much time has passed from the last time it was seen. Additionally, retaining all landmarks in the filter can lead to cumbersome computation times because of an ever growing filter state and the same landmark having multiple representations in the state. When a landmark has not been correlated in $n_{r}$ steps, that landmark is retired. That is, it is removed from the filter state and is no longer tracked. The user-specified parameter, $n_{r}$, can be determined from orbit conditions and the keypoint descriptor method as specified in \cite{dennison_comparing_2021}. 

As landmarks are retired from the filter state, possible duplicate landmarks are removed from the database to maintain a more uniform surface point distribution and avoid excessive memory requirements. When a landmark is retired, all existing retired landmarks within Euclidean distance, $d_r$, of the newly retired landmark's position are found. Then, one-by-one, the newly retired landmark and the existing retired landmarks are compared to see if their covariances overlap significantly via
\begin{equation}\label{eq:eig_overlap}
    l_o = \sqrt{\lambda_*} + \sqrt{\lambda_r} - d_{*,r}.
\end{equation}
Here, $\lambda_*$ and $\lambda_r$ are the maximum eigenvalues of the newly retired landmark and the existing retired landmark covariance, respectively. The Euclidean distance between the two landmarks is $d_{*,r}$. If $l_o < 0$, the landmark with the smaller maximum covariance eigenvalue is retained in the database, and the other landmark is deleted. 

There is one exception that does not follow the retirement process: landmarks initialized via stereovision but never correlated via filter-to-spacecraft correlation within $n_{r}$ steps are deleted from the landmark database altogether. Such landmarks often have relatively large uncertainties and may be the result of an incorrect correlation of two image keypoints.

\section{STATE ESTIMATION FILTER}\label{sec:estimation}
The estimation filter, shown in green in Figure \ref{fig:flowchart}, fuses the pixel measurements from the landmark tracking and stereovision subsystem with the intersatellite RF measurements to estimate the spacecraft states, relative clock offsets, and asteroid characteristics. A UKF\cite{julier_new_1997,thrun_probabilistic_2005} is utilized because it is more accurate for nonlinear systems than an extended Kalman filter and is more computationally efficient than particle filtering and batch estimation techniques\cite{thrun_probabilistic_2005,vallado_fundamentals_2007}. Thus, a UKF provides a good balance between computational cost and accuracy for this application. The ANS UKF computation time is significantly reduced with no loss of accuracy through the recently developed exploiting triangular structure (ETS) technique\cite{stacey_autonomous_2018}. The filter is made robust to system nonlinearities and dynamics modeling deficiencies through measurement underweighting and adaptive estimation of the process noise covariance. 
The state estimation filter is executed entirely on the mothership as illustrated by the green block in Figure \ref{fig:flowchart}.

The estimated filter state is
\begin{equation}\label{eq:state}
    \bm{x} = [\bm{\psi}^T \ \bm{G}^T \ \bm{x}_1^T \ C_{r,1} \ \hdots \ \bm{x}_{n_s}^T \ C_{r,n_s} \ \bm{\delta b}^T \ \bm{L}_1^T \ \hdots \ \bm{L}_{n_l}^T]^T
\end{equation}
where $n_s$ is the number of spacecraft and $n_l$ is the number of landmark states currently in the filter state. Here $\bm{\psi} = [\alpha \ \delta \ \omega]^T$ parameterizes the rotational motion of the asteroid where $\alpha$ and $\delta$ are the right ascension and declination of the asteroid spin axis respectively with respect to the ACI frame, and $\omega$ is the asteroid spin rate\cite{archinal_report_2011}. The current filter implementation assumes the asteroid rotates uniformly about its maximum moment of inertia principle axis, which is true for the majority of asteroids\cite{scheeres_orbital_2016,burns_asteroid_1973}. The vector $\bm{G}$ contains the asteroid gravitational parameter as well as the normalized spherical harmonic coefficients of the asteroid gravity field where the normalization factor is\cite{montenbruck_satellite_2000}
\begin{equation}\label{eq:SH normalization}
    \kappa_{nm} = \sqrt{\frac{(n+m)!}{(2-\delta_{0m})(2n+1)(n-m)!}} \ .
\end{equation}
Here, $n$ and $m$ are the degree and order of the gravity coefficient respectively, and $\delta_{ij}$ is the Kronecker delta function. The inertial Cartesian state and solar radiation pressure coefficient respectively of the $i$th spacecraft are $\bm{x}_i$ and $C_{r,i}$. The clock offset and drift of each deputy with respect to the mothership multiplied by the speed of light are contained in the vector $\bm{\delta b}$. The ACAF 3D position vector of the $i$th tracked landmark is $\bm{L}_i$. An initial mean state estimate and error covariance of the state in Eq. (\ref{eq:state}), except for landmark states, is provided by the brief ground-in-the-loop initialization phase. 

Before each filter measurement update, landmarks are retired (i.e., removed) from the filter state, and new stereovision estimates (see Eq. \ref{eq:ReprojectionGeneral}) from images captured at that epoch are incorporated into the filter state. Landmarks are retired from the filter state when they have not been observed for a specified number of consecutive time steps because SIFT feature matching is not robust to large changes in perspective and illumination conditions. However, the estimates and associated error covariances of landmarks removed from the filter state are maintained in the landmark database as described in Section \ref {sec:ldt} \ref{sec:landmark retirement}.





\subsection{Incorporation of Stereovision Estimates}
Let $\bm{\hat{x}}^-_{k|k-1}$ denote the time updated mean state estimate after retiring landmarks from the filter state but before incorporating new landmarks from stereovision. The associated error covariance is $\mathbf{P}^-_{k|k-1}$. After incorporating a set of $n^+_l$ new landmark position estimates $\hat{\bm{L}} = [\tilde{\bm{L}}_1^{T} \dots \tilde{\bm{L}}_{n_l^+}^{T}]^T$ through stereovision (see Eq. \ref{eq:ReprojectionGeneral}), the mean state estimate is $\bm{\hat{x}}^+_{k|k-1} = [\bm{\hat{x}}^{-T}_{k|k-1} \ \ \hat{\bm{L}}^{T}]^T$ with error covariance
\begin{equation}\label{eq:filter cov sv}
    \mathbf{P}^+_{k|k-1} = 
    \begin{bmatrix}
        \mathbf{P}^-_{k|k-1} &\mathbf{P}_{L,x^-}^T\\
        \mathbf{P}_{L,x^-}  &\mathbf{P}_L
    \end{bmatrix}.
\end{equation} 
The true values of $\bm{\hat{x}}^-_{k|k-1}$, $\bm{\hat{x}}^+_{k|k-1}$, and $\hat{\bm{L}}$ are $\bm{x}^-_{k}$, $\bm{x}^+_{k}$, and $\bm{L}$ respectively. The error covariance of $\hat{\bm{L}}$ is $\mathbf{P}_L$, and $\mathbf{P}_{L,x^-}$ is the cross covariance between the errors of $\hat{\bm{L}}$ and $\bm{\hat{x}}^-_{k|k-1}$. The errors of $\hat{\bm{L}}$ and $\bm{\hat{x}}^-_{k|k-1}$ are correlated because the filter estimates of the spacecraft positions and asteroid rotational parameters are used to compute the stereovision estimates. The computation of $\mathbf{P}_L$ and $\mathbf{P}_{L,x^-}$ through a linearized framework is described in the remainder of this subsection, which is similar to the discussion on consider parameters in \cite{montenbruck_satellite_force_2000}.

After linearizing the argument of the Euclidean norm in Eq. (\ref{eq:ReprojectionGeneral}) for each stereovision estimate about a reference state $\bar{\bm{L}}$, $\hat{\bm{L}}$ can be written as
\begin{align}
    \hat{\bm{L}} 
    &= \bar{\bm{L}} + \underset{\delta \bm{L}}{\text{argmin}} || \mathbf{R}^{-1/2}(\Delta\bm{z} - \mathbf{A}_L \delta \bm{L})||^2\nonumber\\
    &= \bar{\bm{L}} + \mathbf{X}\Delta \bm{z} \label{eq:sv linearization}
\end{align}
Nominally, $\bar{\bm{L}}$ is selected as the most recent solution of Eq. (\ref{eq:ReprojectionGeneral}) obtained through the Gauss-Newton algorithm for each new landmark. Here, $\Delta \bm{z} = \bm{z} - \hat{\bm{z}}$ where $\bm{z}$ is the concatenation of all the recorded pixel measurements used in the computation of $\hat{\bm{L}}$, and $\hat{\bm{z}}$ is the corresponding expected measurements given $\bar{\bm{L}}$ and $\bm{\hat{x}}^-_{k|k-1}$. Specifically, $\bm{z} = \bm{h}(\bm{x}_k^-,\bm{L}) + \bm{\nu}$ and $\hat{\bm{z}} = \bm{h}(\hat{\bm{x}}^-_{k|k-1},\bar{\bm{L}})$ where $\bm{h}$ is the nonlinear measurement model described in Eqs. (\ref{eq:SimpleCameraModel}-\ref{eq:PixelProjection}), and $\bm{\nu}$ is the measurement noise with covariance $\mathbf{R}$. In Eq. (\ref{eq:sv linearization}), $\mathbf{X} = (\mathbf{A}_L^T \mathbf{R}^{-1}\mathbf{A}_L)^{-1} \mathbf{A}_L^T \mathbf{R}^{-1}$ where the Jacobian $\mathbf{A}_L = \frac{\partial \bm{h}}{\partial \bm{L}}$ is evaluated at $\bar{\bm{L}}$ and $\bm{\hat{x}}^-_{k|k-1}$.



Linearizing $\bm{z}$ about $\bar{\bm{L}}$ and $\bm{\hat{x}}^-_{k|k-1}$ yields $\bm{z} = \hat{\bm{z}} + \mathbf{A}_L(\bm{L} - \bar{\bm{L}}) + \mathbf{A}_x(\bm{x}^-_{k} - \bm{\hat{x}}^-_{k|k-1}) + \bm{\nu}$ where the Jacobian $\mathbf{A}_x = \frac{\partial \bm{h}}{\partial \bm{x}^-_k}$ is evaluated at $\bar{\bm{L}}$ and $\bm{\hat{x}}^-_{k|k-1}$. As a result,
\begin{equation}\label{eq:lin dz sv}
    \Delta \bm{z} = \mathbf{A}_L(\bm{L} - \bar{\bm{L}}) + \mathbf{A}_x(\bm{x}^-_{k} - \bm{\hat{x}}^-_{k|k-1}) + \bm{\nu}
\end{equation}
The Jacobians $\mathbf{A}_L$ and $\mathbf{A}_x$ are easily derived from Eqs. (\ref{eq:SimpleCameraModel}-\ref{eq:PixelProjection}) but are not provided here for brevity. Substituting Eq. (\ref{eq:lin dz sv}) into Eq. (\ref{eq:sv linearization}) results in
\begin{equation}\label{eq:Lhat deviation from truth}
    \hat{\bm{L}} = \bm{L} + \mathbf{X}(\mathbf{A}_x(\bm{x}^-_{k} - \bm{\hat{x}}^-_{k|k-1}) + \bm{\nu}).
\end{equation}
Leveraging Eq. (\ref{eq:Lhat deviation from truth}), the terms required in Eq. (\ref{eq:filter cov sv}) are
\begin{align}
    \mathbf{P}_L 
    &=\text{E}[(\bm{L} - \hat{\bm{L}}) (\bm{L} - \hat{\bm{L}})^T]\nonumber\\
    &= \mathbf{X} \mathbf{A}_x \mathbf{P}^-_{k|k-1} \mathbf{A}_x^T \mathbf{X}^T    +   (\mathbf{A}_L^T \mathbf{R}^{-1} \mathbf{A}_L)^{-1}
\end{align}
and
\begin{align}\label{eq:sv cross covariance}
    \mathbf{P}_{L,x^-} 
    &= \text{E}[(\bm{L} - \hat{\bm{L}})  (\bm{x}^-_{k} - \bm{\hat{x}}^-_{k|k-1})^T]\nonumber\\
    &= -\mathbf{X} \mathbf{A}_x \mathbf{P}^-_{k|k-1}
\end{align}

\subsection{Dynamics Modeling}
Each UKF call consists of a time update and a measurement update\cite{thrun_probabilistic_2005}. In both the time and measurement updates, $2n+1$ sigma points are deterministically generated where $n$ is the current number of state variables. In the time update, each sigma point is propagated over the measurement interval. Except for the spacecraft states and $\bm{\delta b}$, all the state variables are treated as constants, and their dynamics are modeled as identity. The spacecraft states are propagated through a fourth-order Runge-Kutta numerical integration where the modeled spacecraft accelerations take into account the spherical harmonic gravity coefficients of the asteroid that are estimated, solar radiation pressure with a constant spacecraft surface area, and third body gravity from the sun.

Although there are various ways to model the many sources of noise that alter the frequency of an oscillator over time, it is common to model an atomic clock as a two state system\cite{galleani_tutorial_2008}. The clock states of the $i$th spacecraft are
\begin{equation}
    \bm{\eta}_i(t) = 
    \begin{bmatrix}
        \eta_i(t)\\
        \dot{\eta}_i(t)
    \end{bmatrix}
\end{equation}
where $i=1$ corresponds to the mothership. The clock offset, which is the difference between the clock time and the true time, is denoted $\eta_i(t)$. The clock drift is $\dot{\eta}_i(t) = \frac{d}{dt} \eta_i(t)$. The dynamical model of the clock states is
\begin{equation}\label{eq:ct clock}
    \dot{\bm{\eta}}_i(t) = 
    \begin{bmatrix}
        0 &1\\
        0 &0
    \end{bmatrix}
    \bm{\eta}_i(t)
    +
    \bm{\epsilon}_{\eta i}(t)
\end{equation}
where $\bm{\epsilon}_{\eta i}(t) \in {\mathbb R}^2$ is a zero-mean white Gaussian process with the power spectral density
\begin{equation}\label{eq:clock psd}
{\mathbf{\widetilde{Q}}}_{\eta i} =
    \begin{bmatrix}
    q_{1_i} &0\\
    0   &q_{2_i}
    \end{bmatrix}.
\end{equation}
The values of $q_1$ and $q_2$ for a specific clock can be estimated by fitting the equation
\begin{equation}\label{eq:fit allan}
    \sigma_y^2(\tau) = \frac{q_1}{\tau} + \frac{q_2 \tau}{3}
\end{equation}
to empirically determined values of the Allan variance $\sigma_y^2(\tau)$ where $\tau$ is the averaging interval\cite{galleani_tutorial_2008,zucca_clock_2005}.

The discrete-time solution to Eq. (\ref{eq:ct clock}) is
\begin{equation}\label{eq:dt clock}
    \bm{\eta}_i(t_k) = {\mathbf{\Phi}}_{\eta}(t_k,t_{k-1})\bm{\eta}_i(t_{k-1}) + \bm{w}_{{\eta i}}(t_k)
\end{equation}
where 
\begin{equation}
    {\mathbf{\Phi}}_{\eta}(t_k,t_{k-1}) =
    \begin{bmatrix}
    1   &\Delta t_k\\
    0   &1
    \end{bmatrix}
\end{equation}
and $\Delta t_k = t_k - t_{k-1}$. It is easily shown that the discrete-time process noise $\bm{w}_{\eta i }(t_k)$ is uncorrelated in time, zero-mean, and normally distributed with covariance \cite{galleani_tutorial_2008}
\begin{equation}\label{eq:clock dt Q}
    {\mathbf{Q}}_{\eta_i}(t_k) = 
    \begin{bmatrix}
    q_{1_i}\Delta t_k+q_{2_i}\frac{\Delta t_k^3}{3}     &q_{2_i}\frac{\Delta t_k^2}{2}\\
    q_{2_i}\frac{\Delta t_k^2}{2}                 &q_{2_i} \Delta t_k
    \end{bmatrix}.
\end{equation}

The biases of the RF measurements are
\begin{equation}\label{eq:define biases}
    \bm{\delta b}_i(t) = 
    \begin{bmatrix}
        \delta b_i(t)\\
        \dot{\delta b}_i(t)
    \end{bmatrix}
    =
    c
    \begin{bmatrix}
        \eta_i(t)-\eta_1(t)\\
        \dot{\eta}_i(t)-\dot{\eta}_1(t)
    \end{bmatrix}
\end{equation}
where $c$ is the speed of light. Note that 
$\bm{\delta b}_1 = \bm{0}$ by definition. Light time delay is neglected in Eq. (\ref{eq:define biases}) because $E[\eta_i(t+\tau)] - \eta_i(t)] = \tau E[\dot{\eta}_i(t)]$ and $\text{Var}(\eta_i(t+\tau) - \eta_i(t)) = q_{1_i}\tau + q_{2_i}\frac{\tau^3}{3}$ are small. The dynamics of $\bm{\delta b}_i(t)$ are described by
\begin{equation}
  \bm{\delta b}_i(t_k) = {\mathbf{\Phi}}_{\eta}(t_k,t_{k-1})\bm{\delta b}_i(t_{k-1}) + \bm{w}_{{\delta bi}}(t_k) 
\end{equation}
where $\bm{w}_{{\delta bi}}(t_k) = c(\bm{w}_{{\eta i}}(t_k) - \bm{w}_{{\eta 1}}(t_k))$.
The vector
\begin{equation}
\bm{\delta b}(t_k) = [\bm{\delta b}_2(t_k)^T \hdots \bm{\delta b}_{n_s}(t_k)^T]^T
\end{equation}
is estimated as part of the filter state (see Eq. (\ref{eq:state})). The associated discrete-time process noise of $\bm{\delta b}(t_k)$ is 
$\bm{w}_{{\delta b}}(t_k) = [\bm{w}_{{\delta b_2}}(t_k)^T \hdots \bm{w}_{{\delta b_{n_s}}}(t_k)^T]^T$.
The random vector $\bm{w}_{{\delta b}}(t_k)$ is normally distributed and uncorrelated in time with covariance
\begin{align}
    {\mathbf{Q}}_{\delta b}&(t_k) = \text{E}[\bm{w}_{{\delta b}}(t_k)\bm{w}_{{\delta b}}(t_k)^T]\nonumber\\ 
        &= 
    \begin{bmatrix}
        {\mathbf{Q}}_{\delta b_{2,2}}(t_k)  &{\mathbf{Q}}_{\delta b_{2,3}}(t_k)  &\hdots  &{\mathbf{Q}}_{\delta b_{2,n_s}}(t_k)\\
        {\mathbf{Q}}_{\delta b_{3,2}}(t_k)  &{\mathbf{Q}}_{\delta b_{3,3}}(t_k)  &\hdots  &{\mathbf{Q}}_{\delta b_{3,n_s}}(t_k)\\
        \vdots  &\vdots    &\ddots     &\vdots\\
        {\mathbf{Q}}_{\delta b_{n_s,2}}(t_k)  &{\mathbf{Q}}_{\delta b_{n_s,3}}(t_k)  &\hdots  &{\mathbf{Q}}_{\delta b_{n_s,n_s}}(t_k)
    \end{bmatrix}.
\end{align}
Assuming the noises of different clocks are uncorrelated,
\begin{align} 
{\mathbf{Q}}_{\delta b_{i,j}}(t_k) &= \text{E}[\bm{w}_{{\delta bi}}(t_k)\bm{w}_{{\delta bj}}(t_k)^T]\nonumber\\
&=
     \begin{cases}
       c^2({\mathbf{Q}}_{\eta_i}(t_k) + {\mathbf{Q}}_{\eta_1}(t_k)) &\ \ \text{for} \ \ \ i = j\\
        c^2 {\mathbf{Q}}_{\eta_1}(t_k) &\ \ \text{otherwise}\\
     \end{cases}
     \label{eq:sol ASNC}
\end{align}

\subsection{Measurement Modeling}
In the measurement update, the predicted measurements for each sigma point are computed. The one-way RF pseudorange measurements between each pair of spacecraft at the epoch $t_k$ are modeled as
\begin{equation}\label{eq:pseudorange meas}
    \rho_{ij}(t_k) = ||\bm{\rho}_{ij}(t_k)|| + \delta b_{j} - \delta b_{i}.
\end{equation}
The transmitting and receiving satellites are indicated by the subscripts $i$ and $j$ respectively. The vector from the $i$th spacecraft antenna to that of the $j$th spacecraft is $\bm{\rho}_{ij}$. The intersatellite one-way Doppler measurements are modeled as
\begin{equation}\label{eq:Doppler meas}
    \dot{\rho}_{ij}(t_k) = \dot{\bm{\rho}}_{ij}(t_k)\cdot\frac{\bm{\rho}_{ij}(t_k)}{||\bm{\rho}_{ij}(t_k)||} + \dot{\delta b}_{j} - \dot{\delta b}_{i}.
\end{equation}
Note that $\delta b_{1} = \dot{\delta b}_{1} = 0$ by definition because the biases are defined relative to the mothership clock. The motion of the spacecraft during the signal time of flight is neglected in Eqs. (\ref{eq:pseudorange meas}) and (\ref{eq:Doppler meas}) because the resulting errors are much smaller than the measurement noise for the considered relative spacecraft motion.

The pixel measurements, $u$ and $v$, of a tracked landmark taken by a camera onboard the $j$th spacecraft are given by the pin-hole camera model in Eqs. (\ref{eq:SimpleCameraModel}-\ref{eq:PixelProjection}). The correspondences between detected 2D image keypoints and landmarks in the filter state are provided by the landmark tracking and stereovision subsystem described in Section \ref{sec:ldt}.

\subsection{Computationally Efficient and Robust Estimation}
In the time update, each sigma point is propagated over the measurement interval using the filter modeled dynamics\cite{thrun_probabilistic_2005}. A traditional UKF would require $(2n+1)n_s$ orbit propagations for each time update where $n$ is the number of state variables and $n_s$ is the number of spacecraft. These orbit propagations dominate the total UKF runtime\cite{stacey_autonomous_2018}. However, the runtime of the ANS UKF is significantly reduced with no loss of accuracy through the recently developed ETS technique\cite{stacey_autonomous_2018}. ETS reduces the number of orbit propagations by exploiting the lower triangular structure of the matrix square root of the filter error covariance. The state ordering in Eq. (\ref{eq:state}) is chosen in order to maximize the computational savings. Although ETS can also be used to reduce the computation time of the measurement update, it is only used here for the time update because the filter computation time is dominated so heavily by the time update.



The performance of a Kalman filter can degrade considerably if there are large, neglected higher order effects. In cases where the state uncertainty is large relative to that of the measurement uncertainty, neglecting significant higher order effects of the nonlinear measurement models can result in a theoretical state error covariance that decreases more quickly than the true state errors, resulting in filter inconsistency \cite{carpenter_navigation_2018}. This phenomenon is especially of concern before filter convergence when the state uncertainty is relatively large. To overcome this obstacle, various measurement underweighting schemes have been proposed to increase the innovation covariance in order to slow the convergence of the filter error covariance. Zanetti gives a comprehensive review of measurement underweighting techniques\cite{carpenter_navigation_2018}. Here, Lear's underweighting method is employed because it is simple and treats the second order effects as a scaling of the first order effects, which adapts the inflation of the innovation covariance according to the size of the state error covariance\cite{carpenter_navigation_2018}.

The performance of a Kalman filter is also heavily dependent on the accuracy of the modeled process noise covariance. Poor modeling of this covariance can lead to filter inconsistency and divergence\cite{schlee_divergence_1967}. Modeling the process noise covariance is especially challenging for an asteroid mission because the dynamical environment is poorly known a priori and the process noise statistics can vary dramatically as the orbit of a satellite changes throughout the mission. In order to make the estimation filter robust to dynamics modeling deficiencies, the recently developed adaptive state noise compensation (ASNC) algorithm is used to efficiently estimate the process noise covariance of the spacecraft states online\cite{stacey_adaptive_2021}. This technique models the process noise covariance of each Cartesian spacecraft state at the $k$th time step as 
\begin{equation}\label{eq:Q cart state small dt}
{\mathbf{Q}}_{k} = 
    \begin{bmatrix}
        \frac{1}{3}\Delta t_k^3 {\mathbf{\widetilde{Q}}}_{k}  &\frac{1}{2}\Delta t_k^2 {\mathbf{\widetilde{Q}}}_{k}\\
        \frac{1}{2}\Delta t_k^2 {\mathbf{\widetilde{Q}}}_{k}   &\Delta t_k {\mathbf{\widetilde{Q}}}_{k}
    \end{bmatrix}
\end{equation}
where $\Delta t_k = t_k - t_{k-1}$. The power spectral density of the unmodeled spacecraft accelerations, ${\mathbf{\widetilde{Q}}} \in {\mathbb{R}}^{3\times 3}$, is assumed diagonal and is adaptively tuned after each filter call. Estimating $\mathbf{\widetilde{Q}}$ is very computationally efficient and can easily incorporate lower and upper bounds on each diagonal element. Each lower bound must be at least zero to ensure ${\mathbf{Q}}_{k}$ is positive semi-definite. 

A coarse upper bound is derived here based on the largest expected unmodeled acceleration the spacecraft will experience, $a_{max}$. Over a small time interval $\Delta t_k$, the greatest amount by which the velocity in any axis is changed due to $a_{max}$ is approximately $a_{max}\Delta t_k$. Assuming that $a_{max}\Delta t_k$ is greater than the change in velocity that will be caused by unmodeled accelerations, $a_{max}\Delta t_k$ is set equal to the square root of an element on the main diagonal of the velocity covariance portion of Eq. (\ref{eq:Q cart state small dt}) to yield the upper bound
\begin{equation}\label{eq:Qtilde J2 upper bound}
    \widetilde{Q}_u = a_{max}^2\Delta t .
\end{equation}
Selection of $a_{max}$ should be based on the modeled dynamics and the best knowledge of the true dynamics. One conservative option is to select $a_{max}$ as the maximum acceleration a spacecraft would experience in its orbit due to the spherical harmonic gravity coefficient $J_2$, which is
\begin{equation}\label{eq:max J2 accel}
    a_{max} = \frac{3\mu J_2 R_{ref}^2}{2r^4}.
\end{equation}
Here, $\mu$ and $R_{ref}$ are the gravitational parameter and reference radius respectively of the central body. The distance of the spacecraft from the central body center of mass is $r$. The value of $a_{max}$ is estimated using the current best knowledge of $\mu$, $J_2$, and $r$. Eq. (\ref{eq:Qtilde J2 upper bound}) can also be used as an initial conservative guess of $\mathbf{\widetilde{Q}}$. 

\section{GLOBAL SHAPE RECONSTRUCTION}\label{sec:shape}
The global shape reconstruction module shown in blue in Figure \ref{fig:flowchart} utilizes a novel technique developed in this section that leverages a priori empirical knowledge of the shape characteristics of celestial bodies. A global asteroid shape model is integral to many mission operations. It helps keep the spacecraft at a safe distance from the asteroid. It also enables the use of a polyhedral gravity model, which is important for operations inside the smallest circumscribing sphere (i.e., Brillouin sphere) of the asteroid where a spherical harmonic gravity model may diverge\cite{werner_exterior_1996}. A spherical harmonic shape model is particularly powerful because it can be paired with a spherical harmonic gravity model to infer characteristics of the body mass distribution \cite{Wieczorek_gravity_2015,besserer_grail_2014}. 

This section describes a novel technique to generate a global spherical harmonic shape model from a 3D point cloud of estimated surface landmark positions. The global shape model is obtained through a regularized least squares fit of the spherical harmonic coefficients to the estimated 3D landmark positions, weighting each landmark according to its uncertainty. A new regularization procedure leveraging generalized cross validation and a priori knowledge of the shape characteristics of celestial bodies is developed to prevent over fitting, thus providing accurate surface reconstruction between landmark point estimates. In ANS, this technique is used periodically to generate a global asteroid 3D shape model based on the landmark point estimates in the landmark database provided by the subsystems discussed in Sections \ref{sec:ldt} and \ref{sec:estimation}. Global shape reconstruction is executed on the mothership as illustrated by the blue block in Figure \ref{fig:flowchart}.


Any surface whose radial distance is a square-integrable function of longitude $\lambda$ and latitude $\phi$ can be represented through a spherical harmonic expansion\cite{Wieczorek_gravity_2015} 
\begin{equation}\label{eq:spherical harmonics expansion}
    r(\lambda,\phi) = \sum_{n=0}^{N} \sum_{m=0}^n (\bar{A}_{nm}\cos{m\lambda} + \bar{B}_{nm}\sin{m\lambda})\bar{P}_{nm}(\sin{\phi}),
\end{equation}
where $r(\lambda,\phi)$ is the distance of the surface from the origin. As $N \to \infty$, Eq. (\ref{eq:spherical harmonics expansion}) describes the surface exactly. In practice, Eq. (\ref{eq:spherical harmonics expansion}) is truncated to some finite maximum degree $N$. The spherical harmonics coefficients $A_{nm}$ and $B_{nm}$ of degree $n$ and order $m$ are normalized such that $\bar{A}_{nm} = \kappa_{nm} A_{nm}$ and $\bar{B}_{nm} = \kappa_{nm} B_{nm}$, where $\kappa_{nm}$ is defined in Eq. (\ref{eq:SH normalization}). The associated Legendre functions $P_{nm}(u)$ are also normalized such that $\bar{P}_{nm}(u) = \kappa_{nm}^{-1} P_{nm}(u)$. For a set of $n_l$ 3D surface points, Eq. (\ref{eq:spherical harmonics expansion}) can be written in matrix form as
\begin{equation}
    \bm{r} = {\matstyle{A}}\bm{s}.
\end{equation}
Here, $\bm{r} = [r_1 \ \dots \ r_{n_l}]^T$
is the concatenation of the Euclidean norm of each landmark position vector. Additionally,
\begin{equation}
    \bm{s} = [\bar{A}_{00} \ \bar{A}_{10} \ \dots \ \bar{A}_{NN} \ \bar{B}_{11} \ \bar{B}_{21} \ \dots \ \bar{B}_{NN}]^T
\end{equation}
is a vector containing the normalized spherical harmonic shape coefficients up through degree and order $N$, and
\begin{equation}
    \matstyle{A} = 
    \begin{bmatrix}
        \bar{P}_{00}(\sin{\phi_1}) &\dots &\bar{P}_{NN}(\sin{\phi_1})\sin{N\lambda_1}\\
         \vdots                      &\ddots  &\vdots\\
         \bar{P}_{00}(\sin{\phi_{n_l}}) &\dots &\bar{P}_{NN}(\sin{\phi_{n_l}})\sin{N\lambda_{n_l}}                          
    \end{bmatrix}
\end{equation}
is defined such that Eq. (\ref{eq:spherical harmonics expansion}) is satisfied for each $r_i$.

Given a set of 3D positions on the surface of an object, spherical harmonic coefficients up to a specified degree and order representing the surface can be obtained through the least squares minimization
\begin{align}\label{eq:ls simple}
    \hat{\bm{s}} 
    & = \underset{\bm{s}}{\text{arg min}} \ 
	||{\matstyle{A}}\bm{s} - \bm{r}||^2\nonumber\\
	& = (\mathbf{A}^T \mathbf{A})^{-1} \mathbf{A}^T \bm{r}.
\end{align}

\noindent Unfortunately, Eq. (\ref{eq:ls simple}) often results in over fitting, which leads to large, false surface protrusions in areas with few landmarks\cite{hajarian_spherical_2015,duxbury_figure_1989}. Over fitting is especially pronounced as the degree of the spherical harmonics expansion increases. To prevent over fitting for weighted least squares problems in general, regularization is often used. One of the most common regularization approaches is referred to as Tikhonov regularization or ridge regression \cite{calvetti_tikhonov_2000,golub_generalized_1979}. 

\subsection{Regularization}
This paper develops a novel approach to prevent over fitting by leveraging a priori knowledge of the shape characteristics of celestial bodies. The root mean square (RMS) power spectrum of the shape spherical harmonic coefficients of a body is defined for each degree $n$ as 
\begin{equation}
    \sigma_{sh}(n) = \left(\frac{1}{2n+1}\sum_{m=0}^n(\bar{A}_{nm}^2 + \bar{B}_{nm}^2)\right)^{1/2},
\end{equation}
which is rotation invariant\cite{Wieczorek_gravity_2015}. Kaula empirically showed that the RMS power spectra of the gravity spherical harmonic coefficients of terrestrial bodies tend to follow a power law\cite{kaula_determination_1963,kaula_theory_2000}, decreasing as $1/n^2$. Similarly, many authors have observed that $\sigma_{sh}(n)$ is also well described by a power law \cite{Wieczorek_gravity_2015,ermakov_power_2018}
\begin{equation}\label{eq:power law}
    \sigma_{sh}(n) = \frac{K}{n^\alpha}.
\end{equation}
for terrestrial and minor bodies. The constants $K$ and $\alpha$ have been empirically determined for various bodies such as Venus\cite{kucinskas_spectral_1993,balmino_spectra_1993}, Mars\cite{balmino_spectra_1993}, and Phobos\cite{gbalmino_gravitational_1994}. For the Earth, the Vening Meinesz rule\cite{vening_remarkable_1951,ermakov_power_2018} defines $\alpha = 2$, although authors with more accurate, modern data estimate $\alpha = ~1.8$\cite{rapp_decay_1989,balmino_spectra_1993}. Recently, Ermakov et al.\cite{ermakov_power_2018} analyzed the data from numerous space missions to various bodies in the solar system and found that $\alpha = 1.67$ and $\alpha = 1.88$ provide a good fit for terrestrial and minor bodies respectively.

This a priori information can be incorporated in the least squares optimization through a Tikhonov regularization term. Additionally taking into account the uncertainty of the estimated landmark positions results in the optimization
\begin{align}\label{eq:ls regul}
    \hat{\bm{s}}
    & = \underset{\bm{s}}{\text{arg min}} \ ||{\matstyle{P}}^{-1/2}({\matstyle{A}}\bm{s} - \bm{r})||^2 + \nu||{\matstyle{\Gamma}}^{1/2} \bm{s}||^2\\
    & = ({\matstyle{A}}^T{\matstyle{P}}^{-1}{\matstyle{A}} + \nu{\matstyle{\Gamma}})^{-1}{\matstyle{A}}^T{\matstyle{P}}^{-1}\bm{r}.
    \label{eq:ls regul solution}
\end{align}
Here $\nu \geq 0$ is a weighting factor, $\matstyle{P}$ is the error covariance of $\bm{r}$, and the matrix square root of ${\matstyle{P}}$ is denoted $\matstyle{P}^{1/2}$. Leveraging the power rule in Eq. (\ref{eq:power law}), the Tikhonov matrix is defined as the diagonal matrix
\begin{align}\label{eq:tikhonov matrix}
    {\matstyle{\Gamma}}^{1/2} &= 
    \begin{bmatrix}
        \text{deg}(s_1)^{\alpha}+\epsilon &0  &\dots  &0\\
        0                   &\text{deg}(s_2)^{\alpha}  &\dots   &0\\
        \vdots              & \vdots               &\ddots &\vdots\\
         0                  &0                      &\hdots  &\text{deg}(s_{n_s})^{\alpha}
    \end{bmatrix}\\
    &=
    \begin{bmatrix}
        \epsilon &0  &\dots  &0\\
        0                   &1  &\dots   &0\\
        \vdots              & \vdots               &\ddots &\vdots\\
         0                  &0                      &\hdots  &N^\alpha
    \end{bmatrix}
\end{align}
where $\text{deg}(s_i)$ is the degree of the $i$th element of $\bm{s}$, and $n_s$ is the cardinality of $\bm{s}$. An arbitrarily small number $\epsilon$ is added to the element in the first row and column of ${\matstyle{\Gamma}}^{1/2}$ in order for ${\matstyle{\Gamma}}^{1/2}$ to be invertible, which is required later to determine $\nu$. The structure of ${\matstyle{\Gamma}}^{1/2}$ encourages surface smoothness similar to other celestial bodies by penalizing large magnitudes for large degree coefficients. The Bayesian interpretation of Eq. (\ref{eq:ls regul}) is that the estimate of the shape coefficients is improved by incorporating the a priori belief that the coefficients are normally distributed with a mean of zero and a covariance of $\frac{1}{\nu}{\matstyle{\Gamma}}^{-1}$. Note that ${\matstyle{\Gamma}}^{-1/2}$ follows the power law in Eq. (\ref{eq:power law}). It is recommended to follow the work of Ermakov et al.\cite{ermakov_power_2018} and use $\alpha = 1.67$ for terrestrial bodies and $\alpha = 1.88$ for minor bodies if no other a priori knowledge of $\alpha$ is available.



The error covariance $\matstyle{P}$ can be approximated through the linear mapping
\begin{equation}\label{eq:lin land covariance}
    {\matstyle{P}} = \frac{\partial \bm{r}}{\partial \bm{x}_L}{\matstyle{P}}_L\frac{\partial \bm{r}}{\partial \bm{x}_L}^T
\end{equation}
where $\bm{x}_L = [\bm{L}_1^T \hdots \bm{L}_{n_l}^T]^T$ is the vector concatenation of the estimated 3D ACAF landmark positions, and ${\matstyle{P}}_L$ is the associated error covariance. The partial derivatives in Eq. (\ref{eq:lin land covariance}) are
\begin{equation}
    \frac{\partial \bm{r}}{\partial \bm{x}_L} = 
    \begin{bmatrix}
        \frac{\partial r_1}{\partial \bm{L_1}} &\bm{0}_{1\times3} &\hdots &\bm{0}_{1\times3}\\
        \bm{0}_{1\times3}  &\frac{\partial r_2}{\partial \bm{L_2}}  &\hdots &\bm{0}_{1\times3}\\
        \vdots  &\vdots  &\ddots    &\vdots\\
        \bm{0}_{1\times3}       &\bm{0}_{1\times3}       &\hdots    &\frac{\partial r_{n_l}}{\partial \bm{L_{n_l}}}
    \end{bmatrix}
\end{equation}
where
\begin{equation}
    \frac{\partial r_i}{\partial \bm{L}_i} = \frac{\bm{L}_{i}^T}{||\bm{L}_i||}.
\end{equation}

\subsection{Selecting the Regularization Weighting Factor}
The weighting factor $\nu$ determines the importance of surface smoothness relative to fitting the data. A variety of methods have been developed to determine the best value of $\nu$ such as the L-curve method\cite{hansen_analysis_1992,hansen_use_1993}, Morozov's discrepancy principle\cite{galatsanos_methods_1992}, minimum $\chi^2$ and equivalent degrees of freedom\cite{thompson_study_1991}, k-fold cross-validation\cite{hastie_elements_2017}, and Allen's predicted residual error sum of squares (PRESS). The PRESS estimate minimizes the average square error between each data point and the corresponding model predicted value when that point is omitted from the regularized least squares solution\cite{golub_generalized_1979}. The generalized cross validation (GCV) estimate is a rotation invariant version of Allen's PRESS estimate\cite{golub_generalized_1979}. Thompson et al.\cite{thompson_study_1991} compare several of these methods and found GCV to generally perform well. Since the GCV estimate tends to work well in practice, it is utilized in this paper \cite{eubank_diagnostics_1985}. To find the GCV optimal $\nu$, Eq. (\ref{eq:ls regul}) is first transformed to the standard form 
\begin{equation}\label{eq:ls regul standard}
    \hat{\bm{s}} = \underset{\bm{s}}{\text{arg min}} \ 
	\frac{1}{n_l}||\bar{\matstyle{A}}\bar{\bm{s}} - \bar{\bm{r}}||^2 + \bar{\nu}||\bar{\bm{s}}||^2
\end{equation}
using the relations
\begin{equation}
\begin{aligned}
    \bar{\matstyle{A}} &= {\matstyle{P}}^{-1/2}{\matstyle{A}}{\matstyle{\Gamma}}^{-1/2}, \hspace{0.5cm}
    &&\bar{\bm{s}} = {\matstyle{\Gamma}}^{1/2}\bm{s} \\
    \bar{\bm{r}} &= {\matstyle{P}}^{-1/2}\bm{r}, &&\bar{\nu} = \frac{1}{n_l} \nu.
\end{aligned}
\end{equation}
Then the GCV optimal $\bar{\nu}$ is the minimizer of the scalar function
\begin{equation}\label{eq:gcv objective}
    V(\bar{\nu}) = \frac{n_l||{\matstyle{B}}\bar{r}||^2}{\Tr({\matstyle{B}})^2}
\end{equation}
where $\Tr(\cdot)$ denotes the trace of a matrix, and ${\matstyle{B}} = {\matstyle{I}} - \bar{\matstyle{A}}(\bar{\matstyle{A}}^T\bar{\matstyle{A}} + n_l\bar{\nu}{\matstyle{I}})^{-1}\bar{\matstyle{A}}^T$. The computation of $V(\bar{\nu})$ for various $\bar{\nu}$ can be reduced from $O(n_s^3)$ to $O(n_s)$ after computing some initial $O(n_s^3)$ matrix decompositions\cite{kent_global_2000}. The denominator of $V(\bar{\nu})$ can also be efficiently and accurately approximated for large scale problems using statistical methods\cite{golub_generalized_1997}.

Most often, $V(\bar{\nu})$ has a single well-defined minimum. However, there can be multiple minima\cite{thompson_cautionary_1989}. The algorithm proposed by Kent and Mohammadzadeh is recommended because it is guaranteed to find the global minimizer of $V(\bar{\nu})$\cite{kent_global_2000}. For simplicity in this paper, the optimal $\bar{\nu}$ is found through a three step process. First, an initial bracket on $\bar{\nu}$ is defined using the range of numerically meaningful values derived by Golub and von Matt\cite{golub_generalized_1997}. Second, a grid search is performed in log space. Third, the $\bar{\nu}$ from the grid search that resulted in the smallest $V(\bar{\nu})$ is refined using the Newton-Rhapson method on the first derivative of $V(\bar{\nu})$. The optimal $\bar{\nu}$ is used to compute $\nu = n_l \bar{\nu}$, which is then used to estimate the shape spherical harmonic coefficients through Eq. (\ref{eq:ls regul solution}). The first and second derivatives of $V(\bar{\nu})$, which are required for the Newton-Rhapson method, are
\begin{align}\label{ea:V 1st der}
    V'(\bar{\nu}) = \ &\frac{n_l}{\Tr({\matstyle{B}})^{2}}\bar{\bm{r}}^T({\matstyle{B}}^T{\matstyle{B}}' + {\matstyle{B}}'^T\matstyle{B})\bar{\bm{r}}\\ 
    &- \frac{2n_l\Tr(\matstyle{B}')}{\Tr(\matstyle{B})^{3}}\bar{\bm{r}}^T\matstyle{B}^T\matstyle{B}\bar{\bm{r}}\nonumber
\end{align}
and
\begin{align}\label{ea:V 2nd der}
    V''(\bar{\nu}) = \ &\frac{n_l}{\Tr(\matstyle{B})^{2}}\bar{\bm{r}}^T(2\matstyle{B}'^T\matstyle{B}' + \matstyle{B}^T\matstyle{B}'' + \matstyle{B}''^T\matstyle{B})\bar{\bm{r}}\\
    &-4\frac{n_l\Tr(\matstyle{B}')}{\Tr(\matstyle{B})^{3}}\bar{\bm{r}}^T(\matstyle{B}^T\matstyle{B}' + \matstyle{B}'^T\matstyle{B})\bar{\bm{r}}\nonumber\\
    &-2n_l\bar{\bm{r}}^T\matstyle{B}^T\matstyle{B}\bar{\bm{r}}\left(\frac{\Tr(\matstyle{B}'')}{\Tr(\matstyle{B})^{3}} - \frac{3\Tr(\matstyle{B}')^2}{\Tr(\matstyle{B})^{4}}\right)\nonumber
\end{align}
where $'$ indicates differentiation with respect to $\bar{\nu}$. Equations (\ref{ea:V 1st der}) and (\ref{ea:V 2nd der}) are used to verify that the obtained $\bar{\nu}$ corresponds to a local minimum. The derivatives of $\matstyle{B}$ with respect to $\bar{\nu}$ are
\begin{align}
    \matstyle{B}' &= n_l\bar{\matstyle{A}}(\bar{\matstyle{A}}^T\bar{\matstyle{A}} + n_l\bar{\nu}\matstyle{I})^{-2}\bar{\matstyle{A}}^T\\
    \matstyle{B}'' &= -2n_l^2\bar{\matstyle{A}}(\bar{\matstyle{A}}^T\bar{\matstyle{A}} + n_l\bar{\nu}\matstyle{I})^{-3}  \bar{\matstyle{A}}^T.
\end{align}

\subsection{Numerical Validation}
\begin{figure}[!th]
\centering
\includegraphics[width=8cm,trim=0 0 0 0,clip]{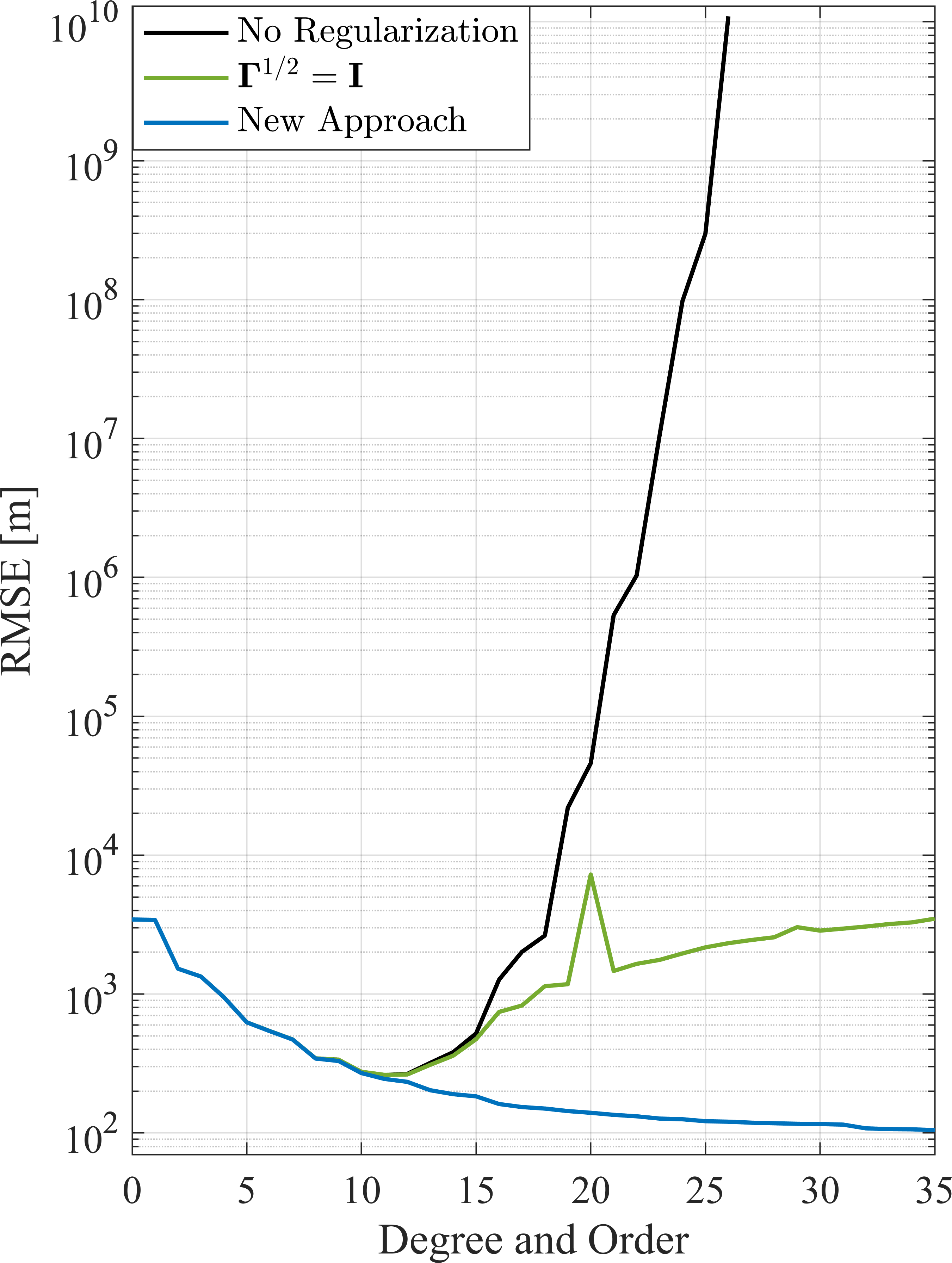}
\caption{RMSE of estimated spherical harmonic shape models of varying degree and order as compared to a triangular mesh model.}
\label{fig:RMSE Eros Different Regularizations}
\end{figure}

\begin{figure*}[!th]
\centering 
\subfigure[\hspace{0.1cm}Triangular mesh]{
\includegraphics[width=.20\linewidth,trim= 0 0 0 0,clip]{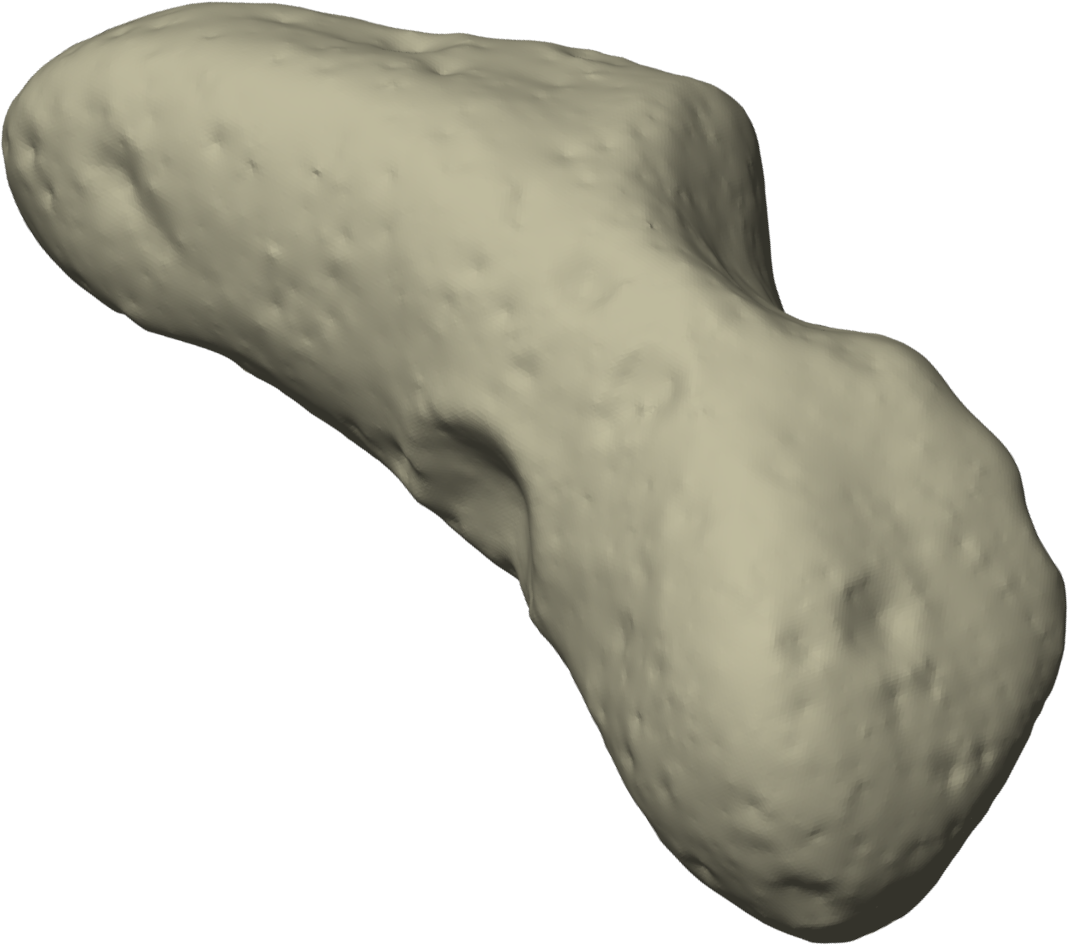}}
\subfigure[\hspace{0.1cm}No regularization]{
\includegraphics[width=.22\linewidth,trim= 0 0 0 0,clip]{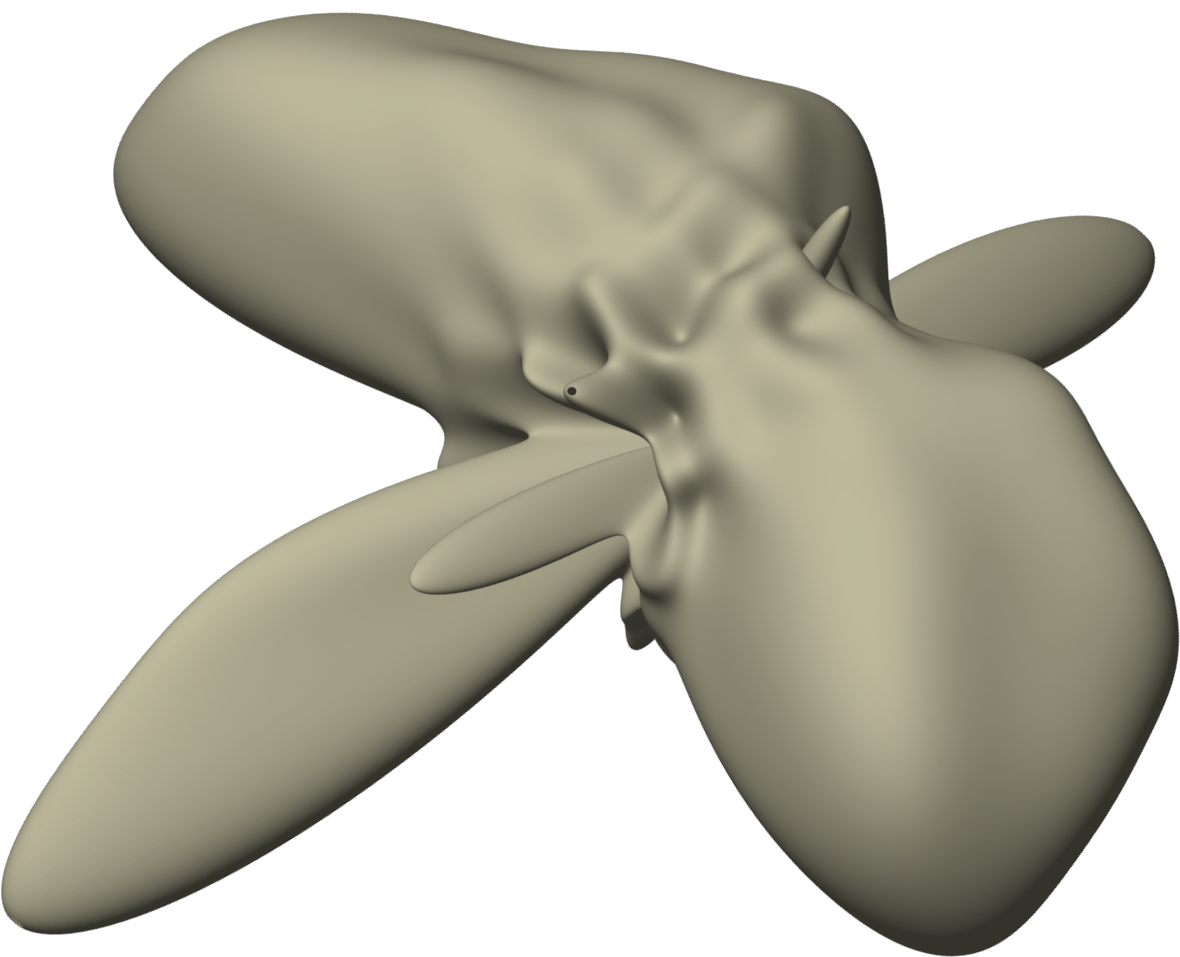}}
\subfigure[\hspace{0.1cm}$\matstyle{\Gamma}^{1/2} = \matstyle{I}$]{
\includegraphics[width=.21\linewidth,trim= 0 0 0 0,clip]{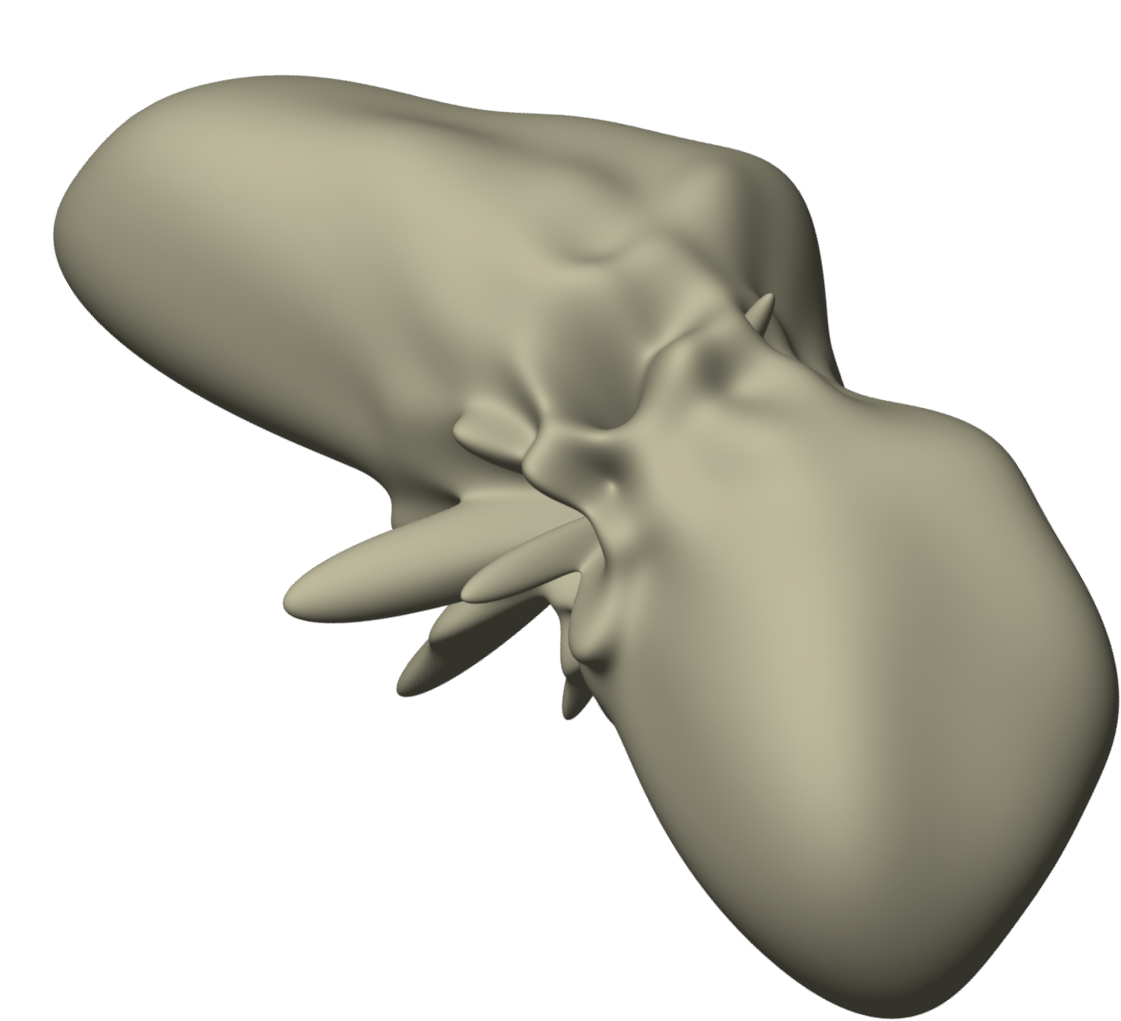}}
\subfigure[\hspace{0.1cm}New approach]{
\includegraphics[width=.20\linewidth,trim= 0 0 0 0,clip]{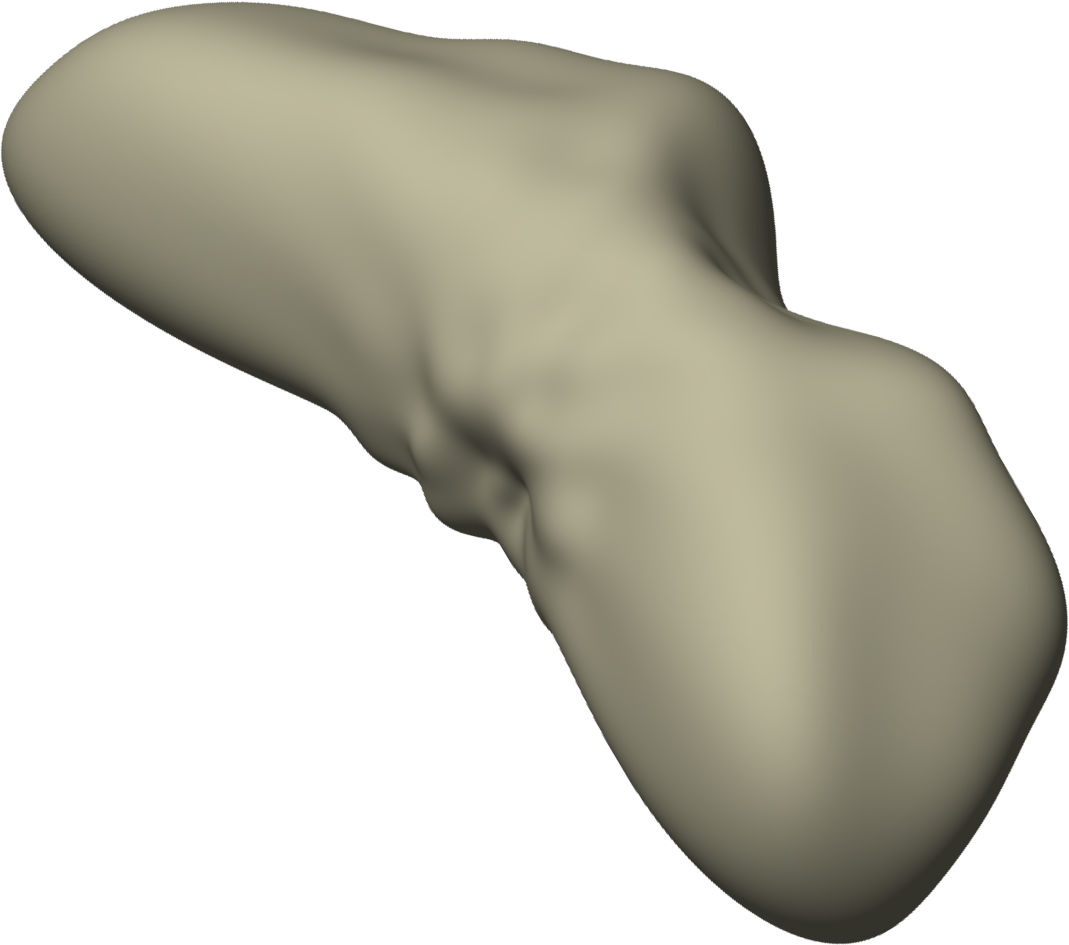}}
\caption{Triangular mesh model compared with estimated degree and order 16 spherical harmonic shape models using 750 vertices randomly sampled from the mesh model.}
\label{fig:shape comparison different regularizations}
\end{figure*}

The benefits of the proposed framework are illustrated in Figures \ref{fig:RMSE Eros Different Regularizations} and \ref{fig:shape comparison different regularizations}. A single set of 750 vertices were randomly sampled from a triangular mesh model of the asteroid Eros that has 200,700 faces\cite{nasa_near_2022}. These sampled points were used to estimate spherical harmonic shape models of varying degree in three ways where all of the vertices were weighted equally. The three approaches used were least squares without regularization as shown in Eq. (\ref{eq:ls simple}) (i.e., $\nu = 0$), the new framework described in this section with the proposed Tikhonov matrix in Eq. (\ref{eq:tikhonov matrix}), and the new framework with a Tikhonov matrix commonly used in regularization, ${\matstyle{\Gamma}}^{1/2} = \matstyle{I}$. Ermakov et al.\cite{ermakov_power_2018} found that $\alpha = 1.84$ best represents minor bodies when excluding Eros from the considered data sets. To simulate a first mission to Eros, $\alpha = 1.84$ is used throughout this paper for the Tikhonov matrix in Eq. (\ref{eq:tikhonov matrix}) although $\alpha = 1.88$ is generally recommended for minor bodies. However, $\alpha = 1.88$ does not result in a significant difference of the estimated coefficients here. The RMS error (RMSE) between the Euclidean norms of the vertices of the triangular mesh model and the corresponding values predicted by the estimated spherical harmonic models is shown in Figure \ref{fig:RMSE Eros Different Regularizations} for each approach. Figure \ref{fig:shape comparison different regularizations} shows the triangular mesh shape model as well as the degree and order 16 spherical harmonic shape models estimated by each of the three approaches. Although all three approaches provide similar results for low degree spherical harmonic models, the proposed approach provides substantially better accuracy for higher degree models by preventing over fitting. Remarkably, the RMSE for the new approach with a degree and order 35 spherical harmonic model is about 2.5 times smaller than the lowest possible RMSE when using either of the other two approaches. The solution of the no regularization approach is not well defined after degree and order 26 because the number of estimated shape coefficients exceeds the number of surface points.

Although the literature reports that GCV tends to work well in practice, there are cases where GCV fails, which are described in depth by Thompson et al\cite{thompson_cautionary_1989}. Of most concern is when the global minimizer of $V(\bar{\nu})$ does not provide a satisfactory result. In these rare cases, GCV tends to dramatically undersmooth, which makes it easy to detect failures\cite{thompson_study_1991}. For the particular application of estimating spherical harmonic shape coefficients, it was found empirically in this work that GCV is likely to fail when the landmarks are concentrated in small regions, there are large regions with relatively few or no landmarks, and a high degree spherical harmonic model is used. Thus, in practice it is recommended to employ at least one other method for choosing $\nu$ as validation. The obtained value of $\nu$ can also be validated by comparing the spectrum of the resulting estimated spherical harmonic coefficients to the a priori spectrum in Eq. (\ref{eq:power law}). If the obtained $\nu$ is unsatisfactory, the degree of the spherical harmonic model should be reduced until a satisfactory $\nu$ is obtained.

\section{CASE STUDY}\label{sec:validation} 
\subsection{Orbit Geometry}
The ANS architecture is validated through a MATLAB simulation of a three-spacecraft swarm orbiting the asteroid 433 Eros. Eros is used as the target asteroid because accurate shape, gravity, and rotational information is available from the NEAR Shoemaker mission\cite{konopliv_global_2002,miller_determination_2002,williams_technical_2002}. The initial swarm mean absolute and relative orbital elements (ROE) are provided in Table \ref{tab:orbit geometry} where the ROE are multiplied by the orbit mean semi-major axis for easy geometrical interpretation\cite{damico_autonomous_2010}. The quasi-nonsingular ROE in Table \ref{tab:orbit geometry} are defined in terms of the classic Keplerian orbital elements as,
\begin{equation} \label{eq:roe}
\begin{bmatrix}
\delta a\\
\delta \lambda\\
\delta e_x\\
\delta e_y\\
\delta i_x\\
\delta i_y
\end{bmatrix}
=
\begin{bmatrix}
(a_d-a_c)/a_c\\
u_d-u_c+(\Omega_d-\Omega_c)\text{cos}(i_c)\\
e_d\text{cos}(w_d)-e_c\text{cos}(w_c)\\
e_d\text{sin}(w_d)-e_c\text{sin}(w_c)\\
i_d-i_c\\
(\Omega_d-\Omega_c)\text{sin}(i_c)
\end{bmatrix}.
\end{equation}

\noindent Here $u=M+w$ is the mean argument of latitude, and the subscripts $c$ and $d$ indicate the chief and deputy respectively. 

\begin{table}[h!]
\caption{Initial swarm mean absolute and relative orbital elements in the ACIC frame. The deputy ROE are defined with respect to the mothership.}
\label{tab:orbit geometry}
\centering
\setlength{\tabcolsep}{2.5pt}
\begin{tabular}{lcccccc}
\hline
    &$a$     &$e$      &$i$  &$\Omega$ &$w$  &$M$\\
Mothership    &45 km   &0.001   &110$^\circ$   &110$^\circ$   &0$^\circ$   &180$^\circ$\\
\hline
&$a\delta a$    &$a\delta \lambda$      &$a\delta e_x$  &$a\delta e_y$ &$a\delta i_x$  &$a\delta i_y$\\
Deputy 1    &0 km   &10 km    &0 km    &0 km   &0 km   &0 km\\
Deputy 2    &0 km   &20 km   &0 km    &0 km   &0 km   &0 km\\
\hline
\end{tabular}
\end{table}

The initial mean semi-major axis of each spacecraft is 45 km, which results in an orbit period of about 25 hrs. The mean orbital elements are transformed to osculating orbital elements through the mapping described by Alfriend\cite{alfriend_spacecraft_2010}, which considers the effects of $J_2$. The mothership orbit is near-circular and slightly retrograde for stability\cite{scheeres_orbital_2016}. The near-polar inclination also provides global coverage of the asteroid surface.

The spacecraft are primarily separated in the along-track direction as described by $\delta \lambda$. The value of $\delta \lambda$ was chosen to balance the need for small intersatellite separation for spacecraft-to-spacecraft feature correlation and large intersatellite separation for passive collision avoidance, stereovision, and observability of the estimated filter state. To prevent the swarm from quickly drifting apart in the along-track direction, the initial mean relative semi-major axis $\delta a$ of each deputy spacecraft relative to the mothership is zero. The details of the reference truth orbit propagations are specified in Table \ref{tab:truth orbit info}. 
 
\begin{table}[!h]
\centering
\caption{Truth orbit propagation parameters.}
\begin{tabular}{l l}
\hline
\bfseries Parameter 		&\bfseries Value\\
\hline
Integration Scheme 			&MATLAB ode45\\
Initial Epoch				&2000 August 1 0 hrs\\ 
Eros Gravity				&Degree and order 15\cite{konopliv_global_2002} \\
Third Body Gravity			&Sun point mass\\ 		
Solar Radiation Pressure	&S/C cross-section normal to\\	
                            &sun, no eclipses\\
\hline
\end{tabular}
\label{tab:truth orbit info}
\end{table}

\begin{figure}[!h]
\centering 
\subfigure[\hspace{0.1cm}Radial-transverse plane]{
\includegraphics[width=5.5cm,trim= 0 0 0 0,clip]{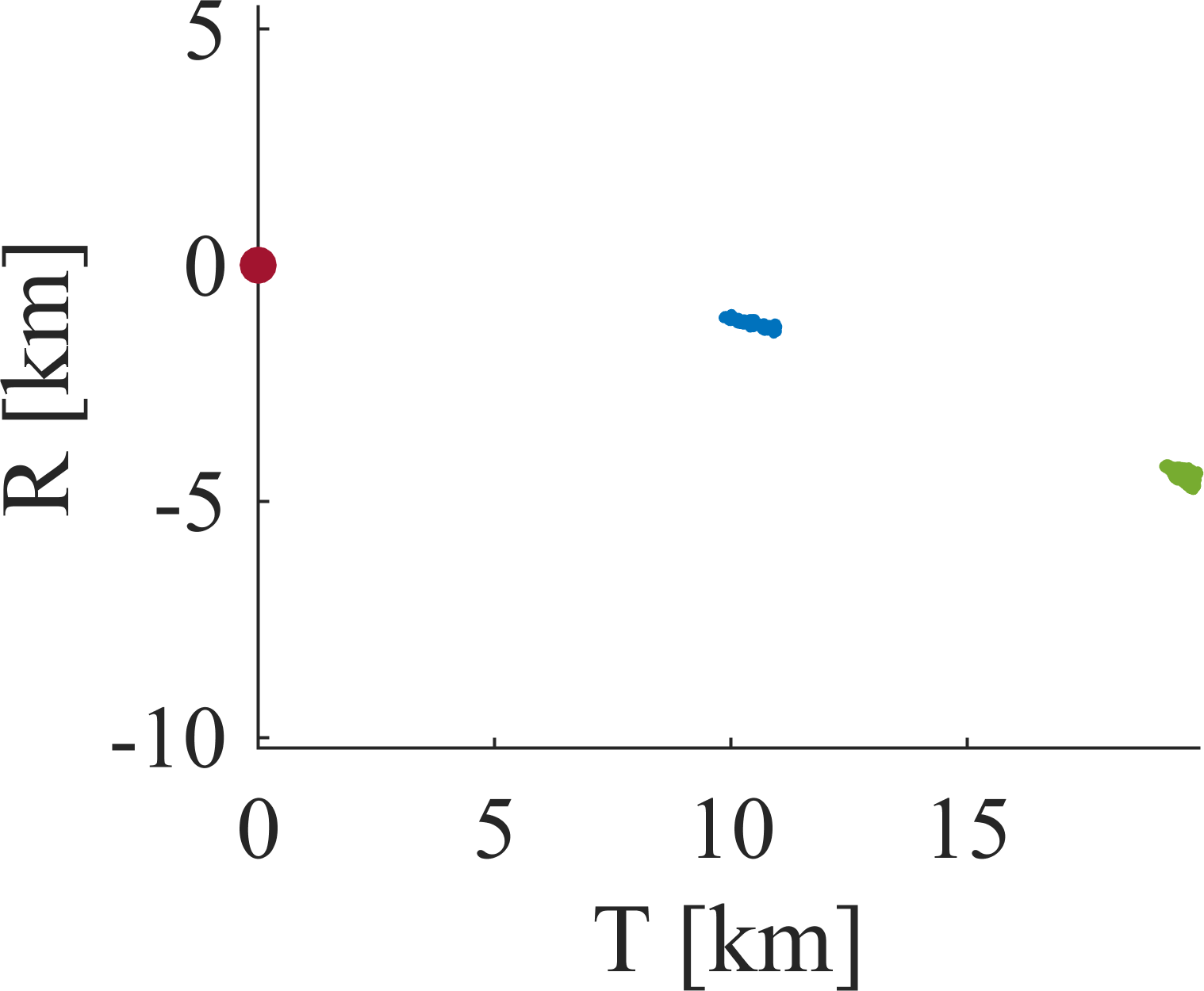}}
\subfigure[\hspace{0.1cm}Radial-normal plane]{
\includegraphics[width=5.5cm,trim= 0 0 0 0,clip]{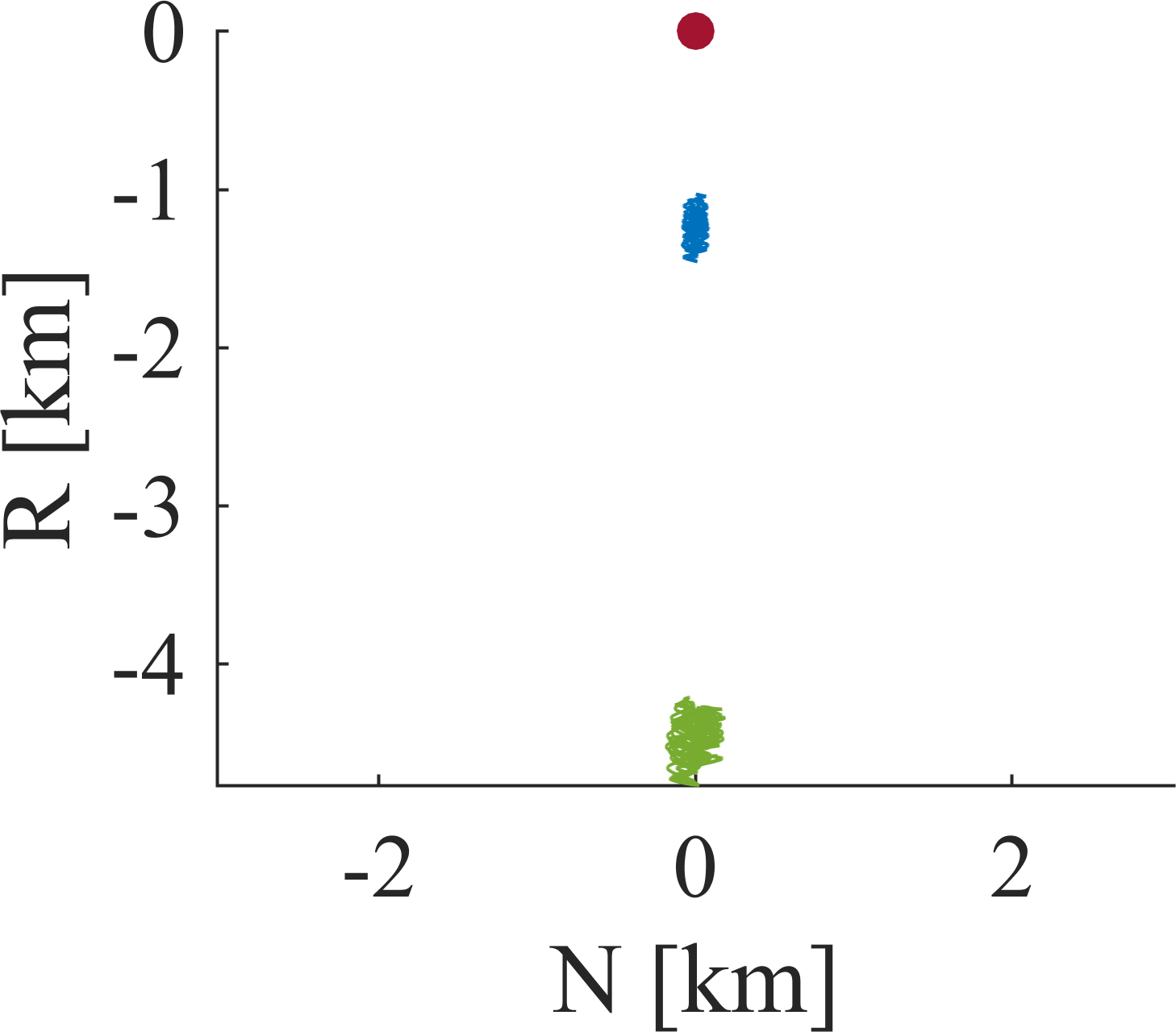}}
\caption{Motion of deputies (blue and green) relative to the mothership (red) in the mothership RTN frame.}
\label{fig:rtn motion}
\end{figure}

The resulting relative motion is shown in Figure \ref{fig:rtn motion}, which demonstrates that safe intersatellite separation is maintained. Occasional station keeping maneuvers would be required to preserve the desired relative orbit geometry. Due to perturbations from gravity coefficients other than $J_2$, the employed mean to osculating transformation can result in a difference between the orbit periods of the mothership and deputies, which creates a secular drift in the spacecraft along-track separation. This phenomenon is exacerbated when decreasing the mean orbit semi-major axis, which would require a mean to osculating transformation that considers more gravity coefficients than $J_2$ in order to accurately match the orbit periods of the spacecraft.

\subsection{Data Generation}
Images and sensor measurements are nominally simulated every 5 min, which results in about 300 measurement intervals over an orbit period. The offset and drift of each spacecraft clock with respect to the true time is stochastically modeled according to Eq. (\ref{eq:dt clock}) with noise parameters that are consistent with the Jackson Labs Low Power Miniature RSR CSAC \cite{jackson_csac_2017}. For this clock, the values of $q_1$ and $q_2$ in Eq. (\ref{eq:clock psd}) are estimated to be $6.2\times 10^{-21}$ s and $1.2 \times 10^{-27}$ s\textsuperscript{-1} respectively using the procedure described in \cite{galleani_tutorial_2008}. The measurements of each spacecraft are recorded when the nominal measurement acquisition time is reached according to the best onboard timing knowledge where the mothership clock is assumed correct. Thus, the true measurement acquisition time of each spacecraft is the desired acquisition time minus the true offset of the onboard clock relative to the true time plus the estimated offset of the onboard clock relative to the mothership clock. 

\begin{figure}[!th]
\centering
\includegraphics[width=8cm]{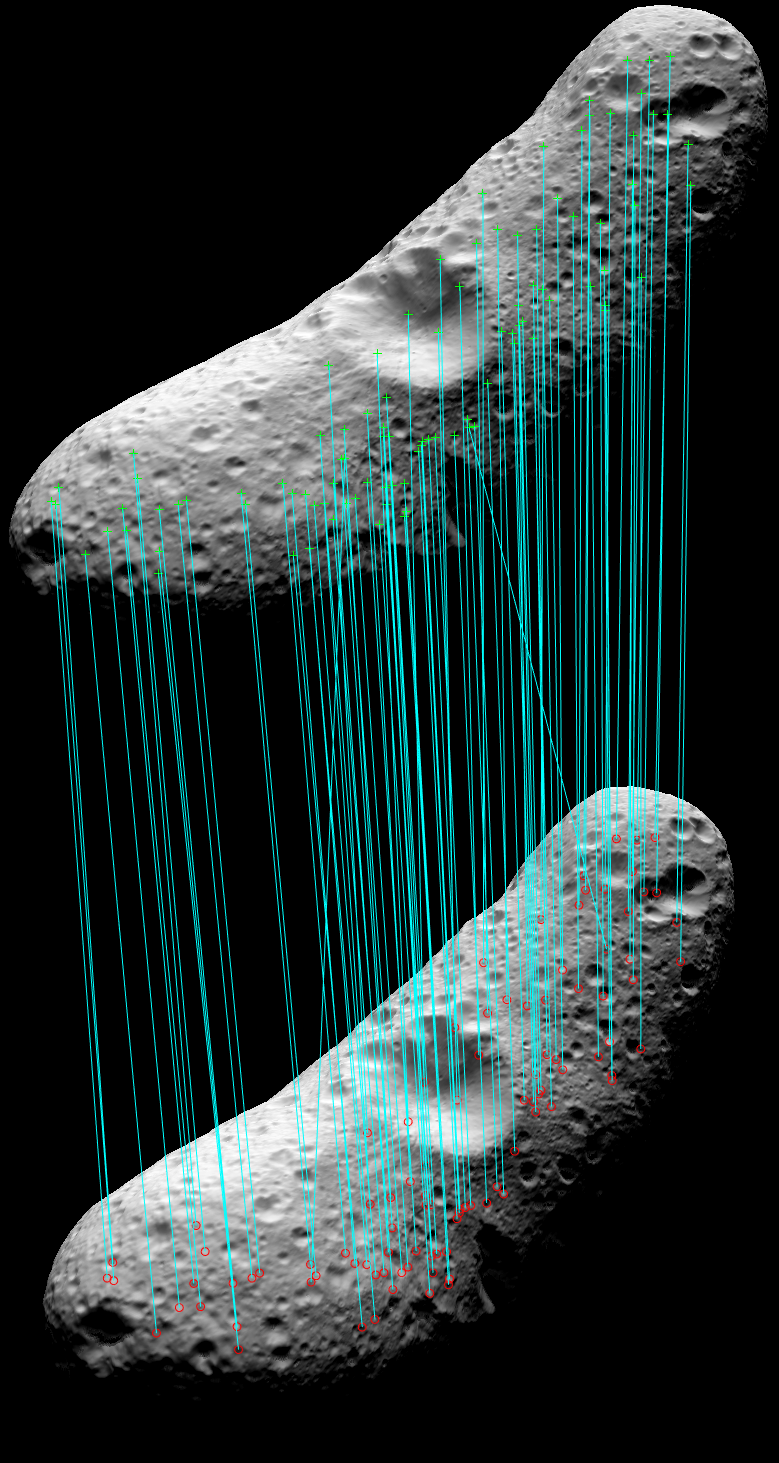}
\caption{Spacecraft-to-spacecraft correlation between two images using SIFT keypoint descriptors, no outlier rejection applied.}
\label{fig:sift_matching}
\end{figure}

Images are generated in MATLAB using OpenGL, the ephemeris of Eros\cite{park_horizons_nodate}, the Eros 3D model from the NEAR Shoemaker mission\cite{konopliv_global_2002}, and the camera model in Eqs. (\ref{eq:SimpleCameraModel} - \ref{eq: def t camera model}). The camera intrinsic parameters are consistent with the GOMSpace NanoCam with an 8 mm lens\cite{GOMSpace2016}. Simulated images incorporate incidence angle shading and soft shadows but no noise or skew \cite{Siegwart2004,heckbert_simulating_nodate}. In the reference truth, the attitude of each spacecraft is defined such that the CF frame z-axis points in the negative radial direction toward the asteroid center of mass, the y-axis is aligned with the orbit angular momentum vector, and the x-axis completes the right handed triad. An example of keypoint correlation between images taken by two different spacecraft at the same nominal epoch is shown in Figure \ref{fig:sift_matching}, which also illustrates the quality of the simulated images.

The intersatellite one-way pseudorange and Doppler measurements are computed according to Eqs. (\ref{eq:pseudorange meas}-\ref{eq:Doppler meas}) and are corrupted with zero-mean white Gaussian noise with standard deviations of 10 cm and 1 mm/s respectively. Error in the star tracker attitude solutions are simulated by multiplying each truth ACI to CF rotation matrix by a stochastic 3-1-2 Euler angle rotation sequence. The angle of each rotation is a Gaussian random variable with a standard deviation of 24 arcsec for the z-axis rotation and 7 arcsec for the x-axis and y-axis rotations, which is consistent with the noise properties of the Blue Canyon Technologies Nano-Star Camera\cite{palo_agile_2013}.

\begin{table}[!h]
\centering
\caption{Landmark tracking and stereovision parameters.}
\begin{tabular}{l l}
\hline

\bfseries Parameter 		&\bfseries Value\\

\hline

Mahalanobis Threshold Probability $(p_m)$   & $0.001$\\
Feature Descriptor Difference $(\delta_{f,t})$       & $100^2$ \\
Mahalanobis Weights $(\hat{w}_{\text{2D}}, \hat{w}_{u}, \hat{w}_{v})$              & $20,\ 5,\ 5$\\
Steps Before Retirement $(n_r)$             & $3$\\
Retired Landmark Euclidean Distance $(d_r)$ & $500$ m \\

\hline
\end{tabular}
\label{tab:lm_parameters}
\end{table}

All user-specified parameters for landmark tracking and stereovision defined in Section \ref{sec:ldt} are summarized in Table \ref{tab:lm_parameters}. 
Image keypoints are detected using the SIFT implementation for MATLAB from VLFeat \cite{vlfeat_vlfeat_nodate}. The probability, $p_m$, used to define the maximum Mahalanobis distance thresholds is set to 0.001, a common choice in hypothesis testing, which results in thresholds of $m_{t,\text{1D}} = 3.291$ and $m_{t,\text{2D}} = 3.717$. The maximum feature descriptor Euclidean distance, $\delta_{f,t}$, used for filter landmark correlation is $100^2$. This value was chosen because landmarks with the most consecutive correlations tended to have descriptor differences between $5^2$ and $90^2$ in preliminary simulations. A landmark is retired from the filter state if it has not been correlated to an image keypoint for $n_r=3$ consecutive time steps, which is nominally 15 min. According to the previous study by the authors \cite{dennison_comparing_2021}, this length of time may be enough for Lowe's keypoint matching to begin to diminish in quality.

The filter a priori state uncertainty is listed in Table \ref{tab:a priori uncertainty and estimation results}. These uncertainties are comparable to the a priori uncertainties typically used in the NEAR orbit determination \cite{williams_technical_2002}. The spacecraft position and velocity a priori uncertainties specified in Table \ref{tab:a priori uncertainty and estimation results} are in each axis. The a priori uncertainty of each normalized spherical harmonic gravity coefficient is 0.005. The error in the initial mean state estimate provided to the filter is randomly sampled according to the initial filter error covariance. The filter modeled dynamics match that of the reference truth except that the filter only estimates the spherical harmonic gravity field up to degree and order eight. Higher order degree gravity coefficients are not explicitly modeled in the filter and are captured through the filter process noise covariance. ASNC is used to estimate the filter process noise covariance of the spacecraft states online through a sliding window of 50 time steps of filter output\cite{stacey_adaptive_2021}. The output of filter calls where no pixel measurements are utilized is excluded from the sliding window due to the poor system observability. Eqs. (\ref{eq:Qtilde J2 upper bound}-\ref{eq:max J2 accel}) are used as an initial guess and upper bound of the unmodeled acceleration power spectral density. The filter measurement models are consistent with the reference truth. Lear's measurement underweighting scheme\cite{carpenter_navigation_2018} is used with a factor of two to increase robustness to measurement nonlinearities as well as errors in matching image keypoints to landmarks in the filter state. The theoretical error covariance of each stereovision estimate is artificially inflated by a factor of two when initially incorporated into the filter error covariance to provide more robustness to potential stereovision biases and outliers. The error standard deviation of each pixel measurement is modeled as 2 px throughout the SNAC pipeline.


\subsection{Results \& Discussion}


The accuracy and precision of the landmark tracking and stereovision module is assessed by ray-tracing image points through the 3D Eros model used to generate the images as described in \cite{dennison_comparing_2021}. For every keypoint detected in an image, its associated ray-traced ACAF position is calculated using the true spacecraft position and attitude as well as the true asteroid state. The correlation algorithms are abbreviated as F2SC (filter-to-spacecraft) and SC2SC (spacecraft-to-spacecraft) in the following figures. 

\begin{figure}[!th]
\centering
\includegraphics[width=8cm,trim=0 0 0 0,clip]{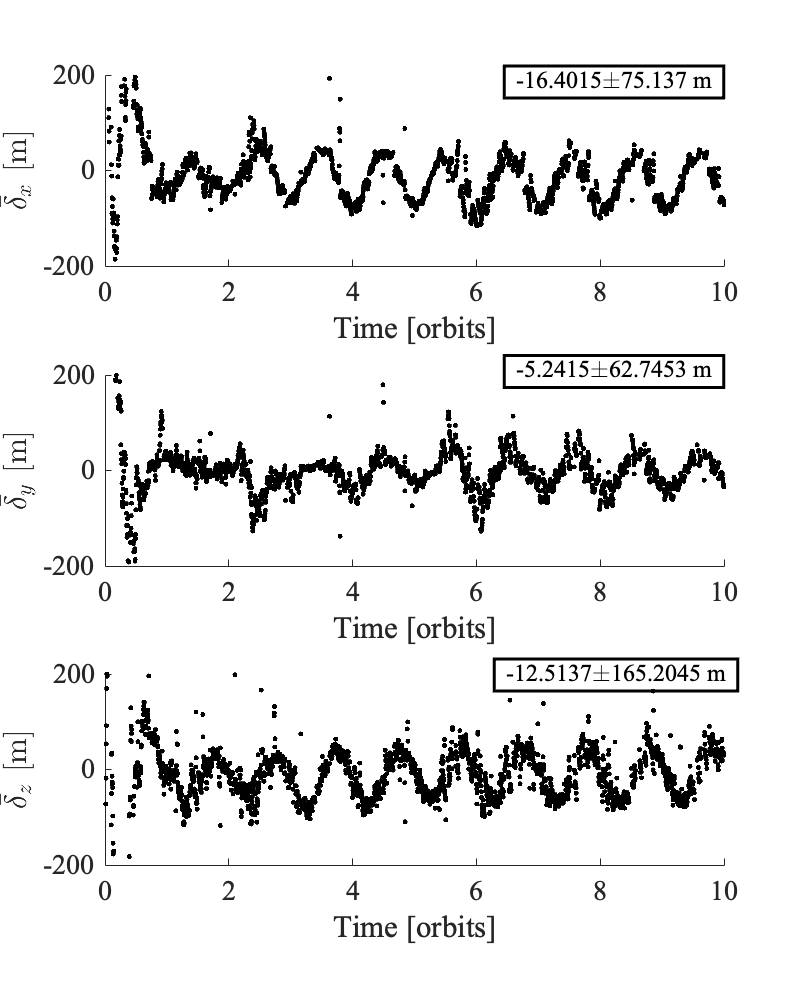}
\caption{Average stereovision error $\bar{\delta}$ at each time step, expressed in the CF frame of the observing spacecraft. Each y-axis is cropped to make trends more visible. The respective simulation mean and standard deviation is in the top right of each subplot.}
\label{fig:sv_error}
\end{figure}

The stereovision error is quantified as the difference between the stereovision-estimated ACAF position and the average of the ray-traced ACAF positions of the keypoints used to generate the stereovision estimate. The ACAF error is then converted to the respective CF frame of each spacecraft used to generate the stereovision estimate. Fig. \ref{fig:sv_error} displays the average CF frame error for all stereovision estimates at each time step. In this figure, the error is quite large in the first 0.5 orbits while the filter state estimates are converging. After that, the error tracks with the spacecraft position error from the UKF. This implies that of all the possible sources of uncertainty in stereovision estimates, spacecraft position error is the largest contributor. The nonzero mean of the stereovision errors can indicate a bias in the estimates. However, the statistical significance of this has not been assessed. The standard deviation of the depth-direction (CF z-axis) is the highest, which is expected because depth is the component that cannot be directly computed from the pixel measurements and requires stereovision.

The accuracy of the correlation algorithms is assessed as the percentage of true positive correlations $p_{\text{TP}}$. The percentage of false positives is $1-p_{\text{TP}}$. A true positive is defined by the Euclidean distance between the ray-traced position of matched landmarks being less than 50 m (0.6\% of Eros' average radius). For spacecraft-to-spacecraft correlation, the ray-traced positions come from the two keypoints in the images. For filter-to-spacecraft correlation, the two ray-traced positions are one, from the keypoint in the current image, and two, the average ray-traced position of the keypoints matched to that landmark in the previous time step. In this simulation, true and false negatives are difficult to assess because it is not guaranteed that SIFT will detect the same keypoints from image to image, even if they do still exist in the image. Because the system is highly flexible to removing obsolete points, false negatives are not concerning while false positives are since false positives can lead the filter and global shape reconstruction astray. 

Table \ref{tab:ldt_stats} shows the correlation algorithm statistics for $p_{\text{TP}}$ and the correlation metrics. The $p_{\text{TP}}$ of each algorithm is exceptionally high, especially when compared to the study on keypoint tracking for Eros by the authors \cite{dennison_comparing_2021}. The increase in true positives in spacecraft-to-spacecraft correlation is attributed to the outlier rejection methods used. Filter-to-spacecraft correlation was aided by the three Mahalanobis distance checks. Without these contributing factors, the performance of both algorithms is expected to be similar to the keypoint descriptor matching results presented in \cite{dennison_comparing_2021} for SIFT. Descriptor differences in table \ref{tab:ldt_stats} are denoted $\Vert \vecstyle{f}_{k1} - \vecstyle{f}_{k2} \Vert^2$ for spacecraft-to-spacecraft and $\Vert \vecstyle{f}_i - \vecstyle{f}_{k^*} \Vert^2$ for filter-to-spacecraft. The average accepted values for descriptor difference and Mahalanobis distance are significantly lower than the thresholds set for them (see Table \ref{tab:lm_parameters}). This indicates that with the chosen thresholds, there is room for error that could be introduced by noisy and skewed images, which were not considered in this paper.

\begin{table}[!h]
\centering
\caption{Simulation mean and standard deviation for spacecraft-to-spacecraft (SC2SC) and filter-to-spacecraft (F2SC) correlation metrics. These are computed from descriptor difference and Mahalanobis distance values of accepted matches.}
\begin{tabular}{lcc}
\hline
\bfseries Parameter &\bfseries Mean   &\bfseries 1-$\bm\sigma$ \\
\hline
$p_{\text{TP},\text{SC2SC}}$ &0.9776 & - \\
$\Vert \vecstyle{f}_{k1} - \vecstyle{f}_{k2} \Vert^2$ &5599.8 &2619.9 \\
$p_{\text{TP},\text{F2SC}}$ &0.9706 & - \\
$\Vert \vecstyle{f}_i - \vecstyle{f}_{k^*} \Vert^2$ &5460.3 &5932.7 \\
$m_{ik^*}$    &0.2063 &0.2436 \\
$m_{ik^*,u}$  &0.1600 &0.2133 \\
$m_{ik^*,v}$  &0.08453 &0.1316 \\
\hline
\end{tabular}
\label{tab:ldt_stats}
\end{table}

\begin{figure}[!th]
\centering
\includegraphics[width=8cm,trim=0 0 0 0,clip]{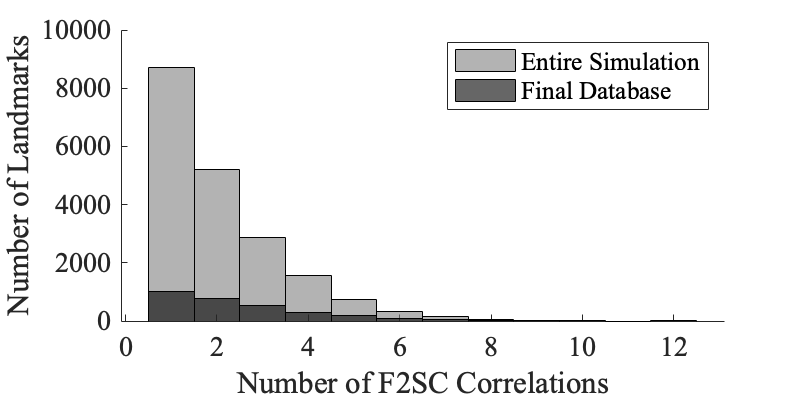}
\caption{Histogram of the number of F2SC correlations of each landmark ever stored in the database throughout the entire simulation and of the landmarks that remained in the final database after ten orbits.}
\label{fig:n_seen}
\end{figure}

The distribution of the number filter-to-spacecraft correlations per landmark is shown in Fig. \ref{fig:n_seen}. The final database has significantly less landmarks than were seen throughout the entire simulation. However, a larger proportion of the total landmarks in the final database had more than one correlation. This could indicate that many landmarks were initialized multiple times or that only a portion of the SIFT keypoints were suitable to tracking sequentially under the lighting and perspective change conditions. This has two implications. One, the landmark retirement process eliminated these spurious landmarks and kept the database from becoming unnecessarily large, which reduces memory requirements. Two, not being able to re-correlate landmarks after retirement leads to re-initialization and the majority of landmarks having three or less measurements over their lifespan. This is not enough measurements for the UKF to significantly improve upon the landmark stereovision position estimates. A method for re-correlation after retirement could greatly reduce the final landmark position errors.

\begin{figure*}[!th]
\centering 
\subfigure[\hspace{0.1cm}True Shape Model]{
\includegraphics[width=.28\linewidth,trim= 0 0 0 0,clip]{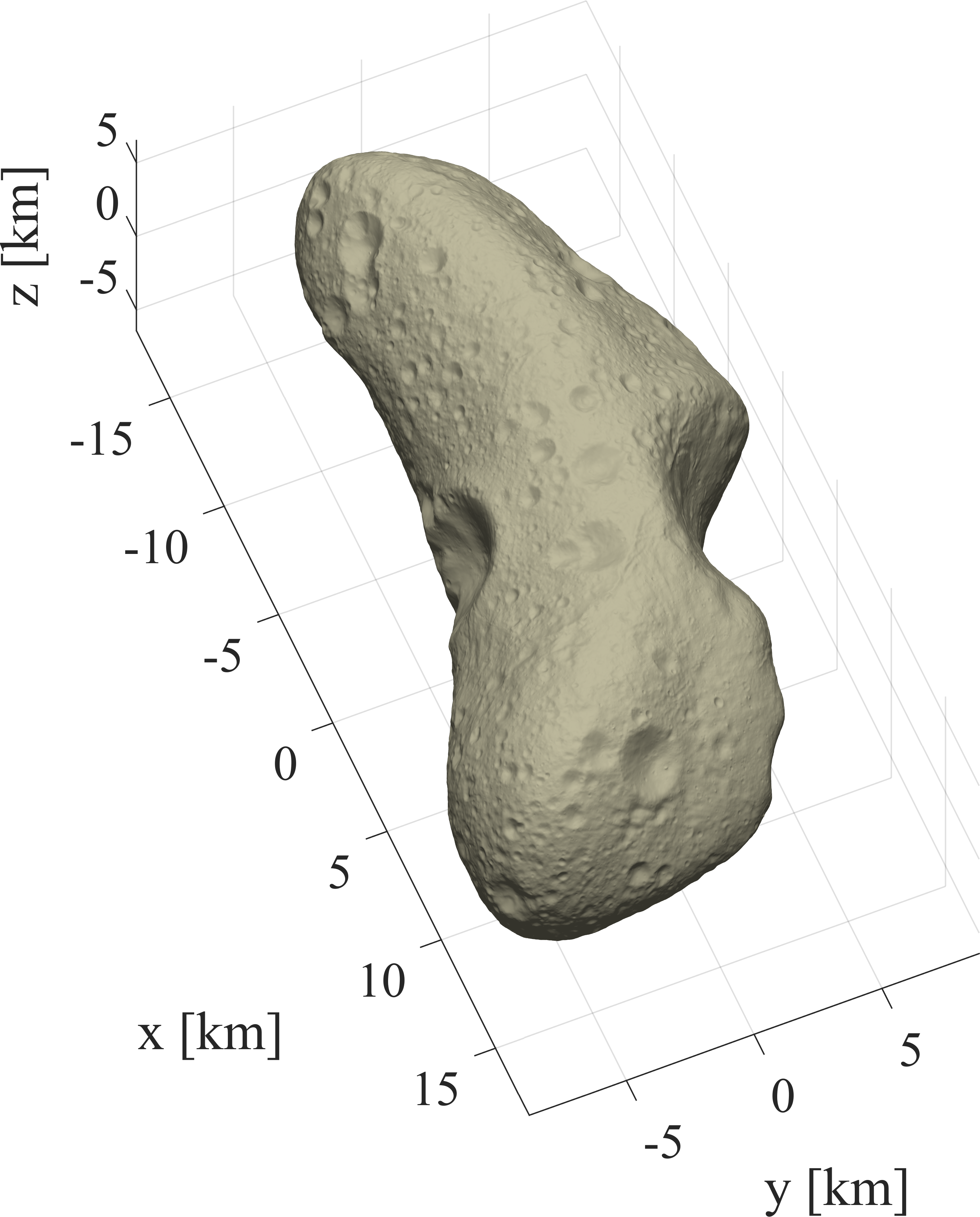}}
\subfigure[\hspace{0.1cm}Landmark Database]{\label{fig:final database}
\includegraphics[width=.28\linewidth,trim= 0 0 0 0,clip]{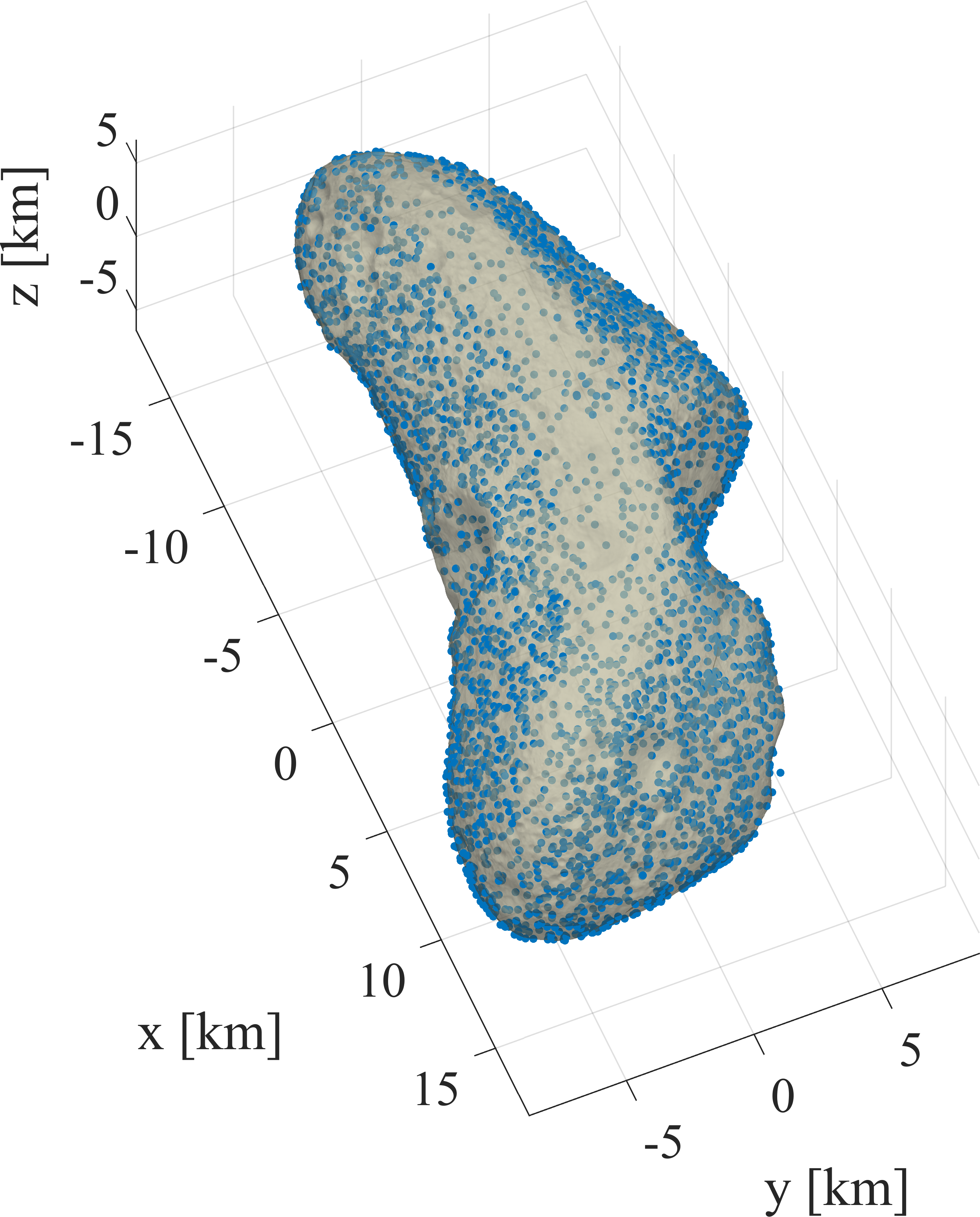}}
\subfigure[\hspace{0.1cm}Global Shape Reconstruction]{
\includegraphics[width=.28\linewidth,trim= 0 0 0 0,clip]{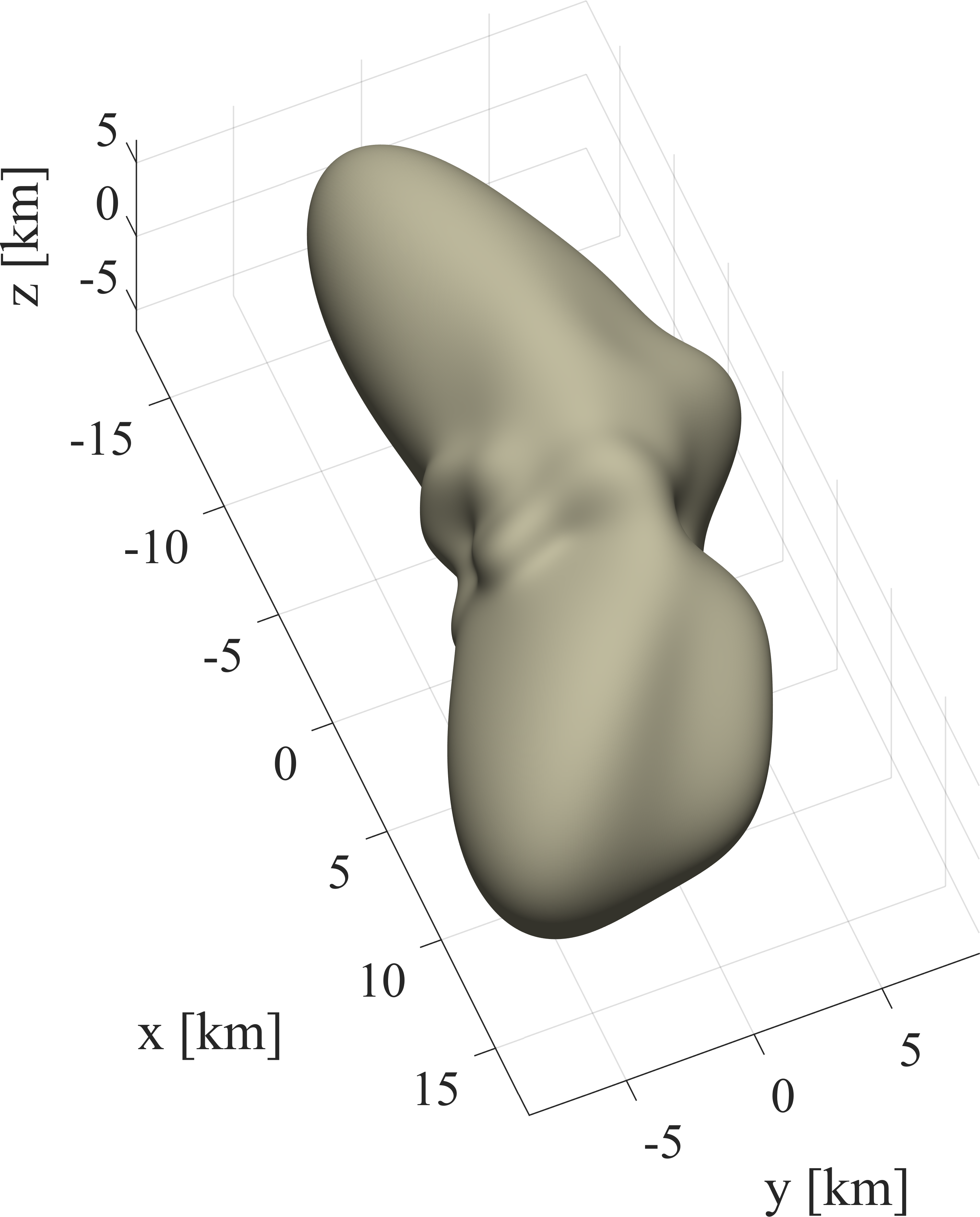}}
\caption{Comparison of a) the true asteroid shape model used to generate images in the simulation, b) the final landmark database plotted with the true shape model made transparent, and c) the estimated degree and order ten spherical harmonic shape model.}
\label{fig:3D shape from SIFT}
\end{figure*}

The final landmark database consisted of 3059 landmarks and is displayed in Figure \ref{fig:final database}. There is a relatively even distribution of landmarks across the surface of Eros with few outliers. However, there is a patch on the +Z surface with no landmarks. This is due to an adverse lighting condition where the average angle between the Sun and the +Z ACAF axis was $105.2^\circ$, meaning the patch was typically on the edge of the shadow. The otherwise uniform distribution of the final database shows that the landmark tracking and stereovision module and state estimation module worked well together to refine the majority of the stereovision estimates to the surface of Eros. 

\begin{figure}[!th]
\centering 
\includegraphics[width=8cm]{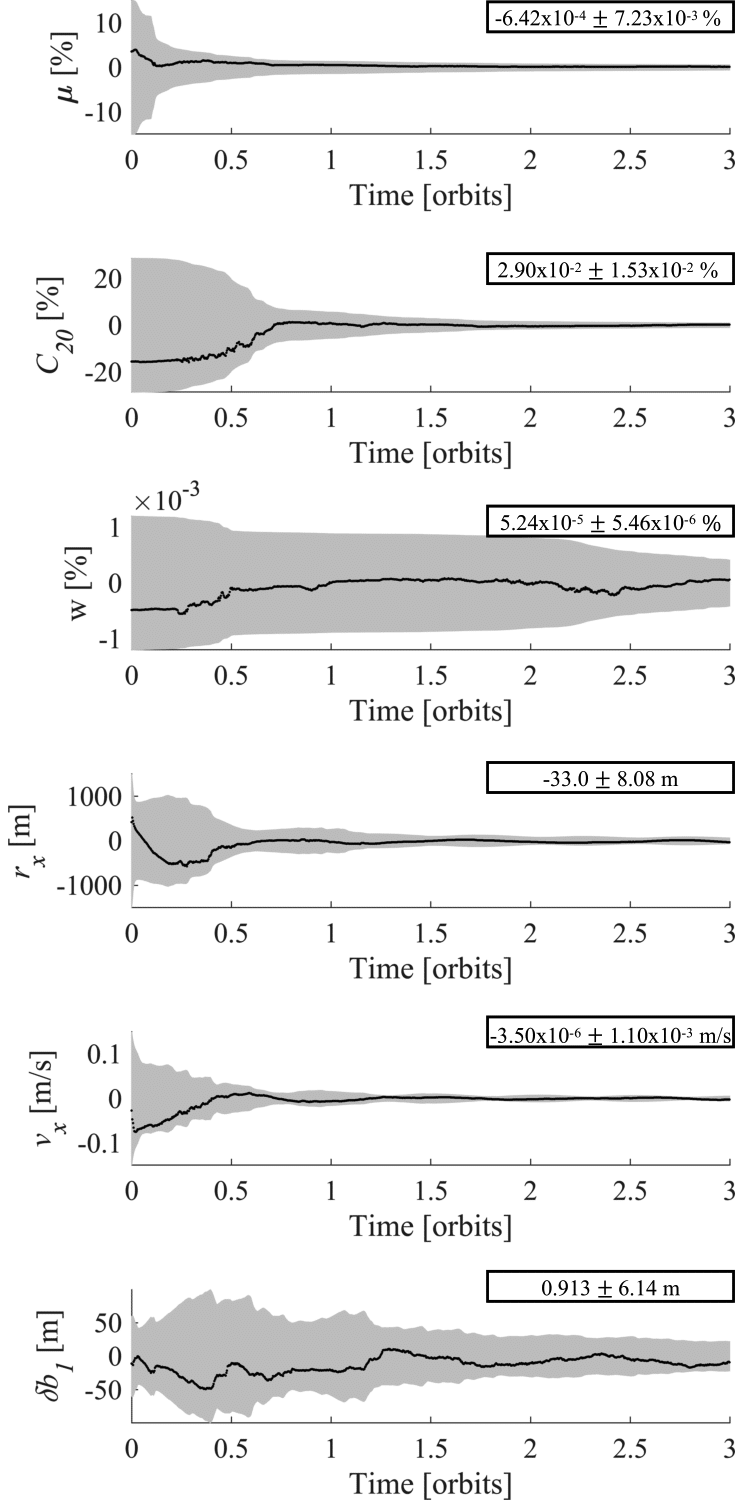}
\caption{Filter estimation error and associated 3-$\sigma$ formal uncertainty bound over first three orbits. From top to bottom, considered state parameters are asteroid gravitational parameter, $C_{2,0}$ spherical harmonic gravity coefficient, and spin rate followed by x-components of the chief position and velocity vectors and offset of first deputy clock relative to the mothership clock multiplied by the speed of light. Boxed numbers are error mean and standard deviation over last orbit of simulation.
}
\label{fig:filter convergence}
\end{figure}

Filter performance is shown in Figure \ref{fig:filter convergence} and Table \ref{tab:a priori uncertainty and estimation results}. Figure \ref{fig:filter convergence} shows qualitative filter convergence behavior for several state parameters. The filter initially converges very rapidly but remains consistent. Table \ref{tab:a priori uncertainty and estimation results} shows quantitative filter performance using two error metrics. The first metric is the RMSE where the error is the difference between the true and estimated state parameters. In computing the RMSE of the specified parameter, all the errors of each state element associated with that parameter are considered over the final orbit of simulation. For example, the provided spacecraft position RMSE considers the position errors in each axis for each spacecraft. The second performance metric in Table \ref{tab:a priori uncertainty and estimation results}, denoted $\bar{\sigma}$, is the square root of the mean variance provided by the filter. Similar to the RMSE, $\bar{\sigma}$ considers the variances of each state element associated with that parameter over the final orbit of simulation.

\begin{figure}[!th]
\centering
\includegraphics[width=8cm,trim=0 0 0 0,clip]{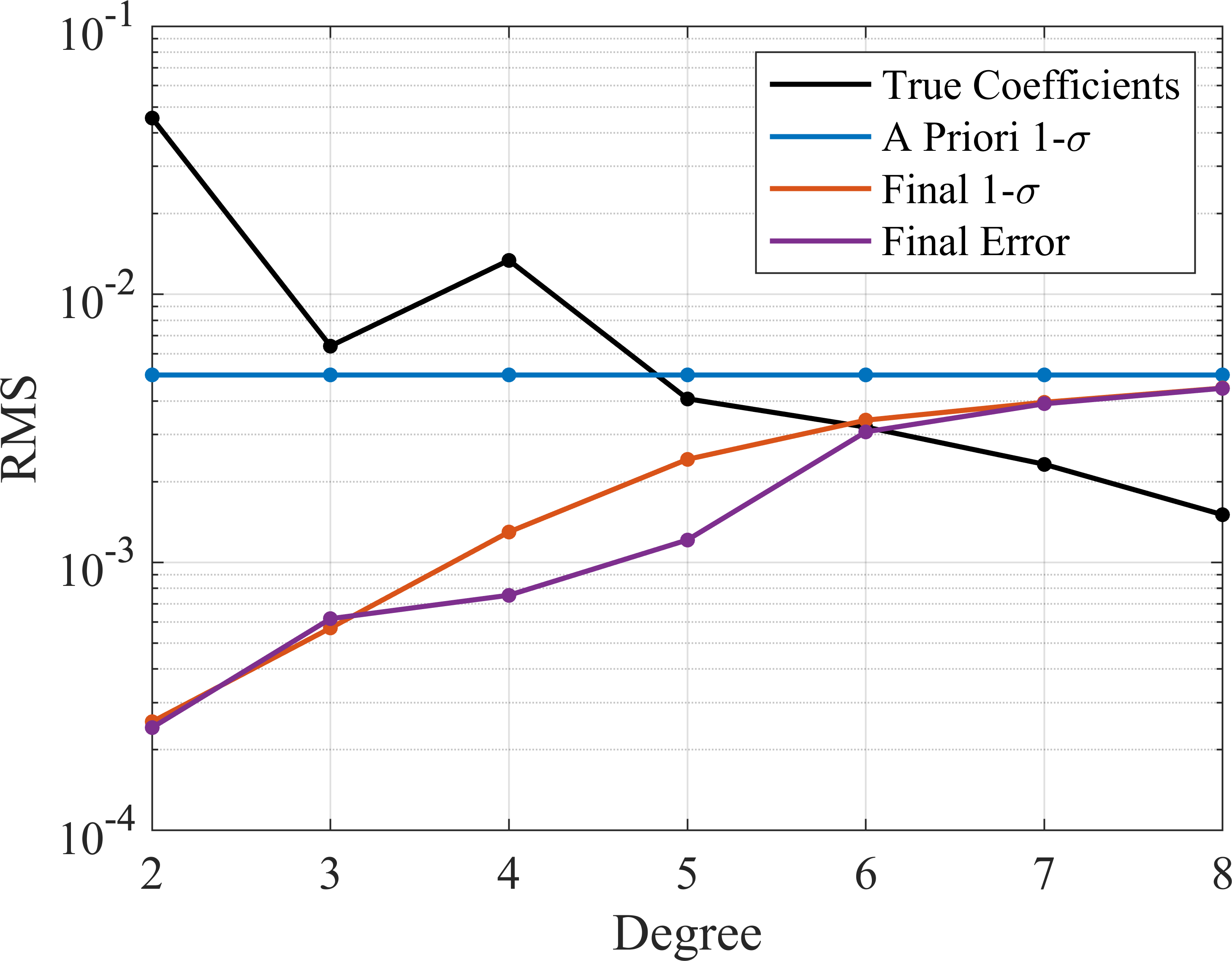}
\caption{Comparison of the RMS of each degree of the true spherical harmonic gravity coefficients, the 1$-\sigma$ a priori uncertainty provided to the filter, and the estimation error and 1$-\sigma$ uncertainty at the end of the simulated case study.}
\label{fig:RMS Gravity}
\end{figure}

The RMSE and $\bar{\sigma}$ in Table \ref{tab:a priori uncertainty and estimation results} are similar for each parameter. The fact that the filter-provided formal uncertainties are consistent with the true errors indicates the filter is functioning well. Furthermore, the a priori uncertainty of each parameter is significantly reduced by the end of the simulation. In particular, the uncertainty of the asteroid gravitational parameter is reduced by nearly two orders of magnitude. The accuracy in estimating the spherical harmonic gravity field is illustrated in Figure \ref{fig:RMS Gravity}. This figure shows that the asteroid gravity spherical harmonic coefficients were estimated to an uncertainty approximately equal to or less than the magnitude of the true coefficients up through degree and order six. Again, the estimation errors are consistent with the filter formal uncertainties. 

As shown in Figure \ref{fig:3D shape from SIFT}, a degree and order ten shape model was fit to the final landmark database using the new approach in Section \ref{sec:shape}. The RMSE between the Euclidean norms of the vertices of the truth shape model and the corresponding values predicted by the estimated spherical harmonic model is 330 m, which is 3.9\% of the true asteroid average radius\cite{giorgini}. The absence of landmarks on the +Z axis side of the asteroid lead to a small, false ridge there in the reconstructed shape. Considering the lack of landmarks in this region, the achieved shape reconstruction accuracy is remarkable. These results demonstrate the robustness of the proposed SNAC framework to adverse lighting conditions. More accurate shape reconstruction can likely be achieved through additional orbits of observations, especially if lighting conditions enable landmarks to be tracked on the +Z axis side of the asteroid. 


\begin{table}[!h]
\centering
\caption{Comparison of the a priori filter state uncertainty with the estimation performance over the final orbit of simulation. Percentages refer to percent of the true value.}
\begin{tabular}{llccc}
\hline
&\bfseries Parameter &\bfseries A Priori   &\bfseries Filter &\bfseries Filter\\
&                    &1-$\sigma$        &RMSE        &$\bar{\sigma}$\\
\hline
\parbox[t]{2mm}{\multirow{3}{*}{\rotatebox[origin=c]{90}{\bfseries Asteroid \hspace{0.1cm}}}} 
&$\mu$ [\%]           &5                 &7.24 $\times 10^{-3}$ &7.05 $\times 10^{-2}$\\
&$\alpha$ [$^\circ$]  &0.1               &1.57 $\times 10^{-2}$ &1.34 $\times 10^{-2}$\\
&$\delta$ [$^\circ$]  &0.1               &4.78 $\times 10^{-3}$ &1.25 $\times 10^{-2}$\\
&$w$ [\%]             &4$\times 10^{-4}$ &5.27 $\times 10^{-5}$ &2.19 $\times 10^{-5}$\\
\hline
\parbox[t]{2mm}{\multirow{3}{*}{\rotatebox[origin=c]{90}{\bfseries Spacecraft \hspace{0.1cm}}}} 
&Position [m]       &500      &26.8     &8.01\\
&Velocity [mm/s]    &50       &1.24     &0.522\\
&$C_R$ [\%]         &10       &8.90     &4.69\\
&$b_R$ [m]          &20       &8.36     &3.18\\
&$b_D$ [mm/s]       &2        &0.624    &0.471\\
\hline
\end{tabular}
\label{tab:a priori uncertainty and estimation results}
\end{table}

Table \ref{tab:computation} summarizes the computation times of the SNAC pipeline for a MATLAB implementation on a 3.6 GHz Intel Core i7-9700K CPU. All the computations listed in  Table \ref{tab:computation} were completed for each measurement interval except for shape reconstruction, which was only completed once at the end of the simulation. In a real mission, all the computations in Table \ref{tab:computation} would only be completed onboard the mothership except for SIFT detection, which would be completed by each spacecraft for each image it records. Other than implementing ETS to reduce the filter computation time, little effort was made to optimize the computational efficiency of the SNAC framework. Thus, these computation times represent upper bounds that can likely be reduced significantly. Additionally, the computation time of the landmark tracking and stereovision module can be reduced by implementing a more efficient matching search algorithm like k-d trees \cite{bentley1975multidimensional} and by using features that are detected more efficiently such as SURF \cite{bay_surf_2006} and ORB \cite{rublee_orb_2011}. ETS provides a greater percent reduction in filter computation time when more landmarks are included in the estimated state. For example, when there are no landmarks or 150 landmarks in the filter state, ETS reduces the filter computation time by about 11\% and 83\% respectively as compared to a traditional UKF.

\begin{table}[!h]
\centering
\caption{Average and worst-case computation times for the various algorithms within the ANS SNAC framework.}
\begin{tabular}{lcc}
\hline
\bfseries Algorithm &\bfseries Average [s]   &\bfseries Worst Case [s] \\
\hline
SIFT (Single Image) &0.571 &0.686 \\
SC2SC Correlation &7.27$\times10^{-3}$ &2.25$\times10^{-2}$ \\
Stereovision &5.81$\times10^{-3}$ &2.37$\times10^{-2}$ \\
F2SC Correlation &0.136 &0.570 \\
State Estimation Filter &0.820 &1.175 \\ 
Shape Reconstruction &10.8 &- \\
\hline
\end{tabular}
\label{tab:computation}
\end{table}


\section{CONCLUSIONS}\label{sec:conclusion} 
Completed asteroid missions have relied heavily on human oversight and ground-based resources such as the NASA Deep Space Network. In contrast, this paper develops an autonomous asteroid characterization algorithmic framework for the Autonomous Nanosatellite Swarming (ANS) mission concept. Specifically, the novel ANS simultaneous navigation and characterization (SNAC) architecture is addressed. SNAC is a new class of estimation problem defined in this paper as a superset of simultaneous localization and mapping. For ANS, SNAC includes estimating the spacecraft states as well as the asteroid gravity field, rotational motion, and shape. Through autonomy and small satellite technology, ANS improves performance and reduces mission cost, which could enable a greater number of future asteroid missions. 

ANS comprises a mothership and one or more smaller nanosatellites in closed orbits equipped with low size, weight, power, and cost avionics. Stereovision and optical feature tracking algorithms are utilized in a novel manner to provide initial estimates of asteroid landmark positions and optical pixel measurements of these landmarks over time. The optical landmark measurements are fused with intersatellite radio-frequency measurements in a computationally efficient and robust unscented Kalman filter to simultaneously estimate the spacecraft states and the asteroid gravity field, rotational motion, and landmark positions. The estimated landmark positions are utilized to reconstruct a global spherical harmonic asteroid shape model through a new technique that leverages a priori knowledge of the shape characteristics of small celestial bodies.

High-fidelity numerical simulations of three spacecraft orbiting the asteroid 433 Eros demonstrate that ANS achieves robust and accurate SNAC. The proposed approach does not require an a priori shape model and uses only low size, weight, power, and cost avionics. The designed orbit geometry avoids collisions throughout the simulation and results in intersatellite separations that are small enough for SIFT feature correlation between spacecraft and large enough for accurate stereovision. The stereovision errors in the depth direction are generally less than 6\% of the average asteroid radius. Landmark tracking errors are less than 3\% of the average asteroid radius in the non-depth directions. After just ten orbits, the uncertainty in every filter state parameter is reduced significantly. Specifically, the asteroid gravity spherical harmonic coefficients were estimated to an uncertainty approximately equal to or less than the magnitude of the true coefficients up through degree and order six. Furthermore, the novel spherical harmonic global shape estimation technique was able to reconstruct the shape of Eros with an RMSE of 3.9\% of the average asteroid radius.

\bibliographystyle{IEEEtaes}
\bibliography{main-double.bib}

\begin{IEEEbiography}[{\includegraphics[width=1in,height=1.25in,clip,keepaspectratio]{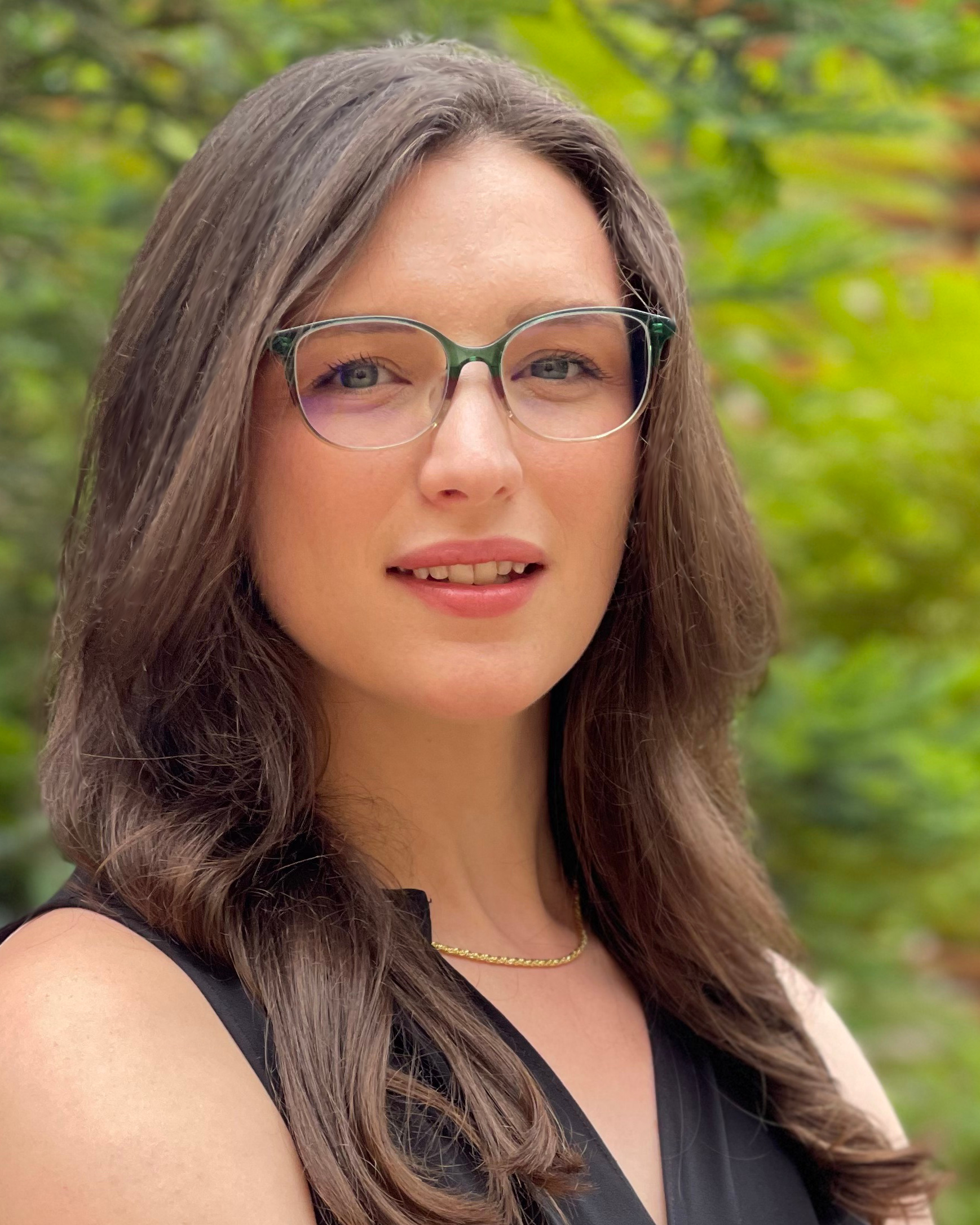}}]{Kaitlin Dennison}{\space} earned her B.S. in mechanical engineering from the University of Connecticut in Storrs, CT (2017). She received her M.S. in aeronautics \& astronautics from Stanford University in Stanford, CA (2019) where she is currently pursuing her Ph.D. in aeronautics \& astronautics. 

Kaitlin worked with the Lawrence Livermore National Laboratory on telescope optics to aid the search for exoplanets. She was also a scholar for the Air Force Research Laboratory where she improved the spacecraft tracking algorithms involving telescope imagery. Additionally, she interned for Blue Origin where she progressed LIDAR-based navigation methods. Her dissertation research in the Space Rendezvous Laboratory advances multi-agent optical tracking and structure from motion in spacecraft swarms with limited resources.
\end{IEEEbiography}




\begin{IEEEbiography}[{\includegraphics[width=1in,height=1.25in,clip,keepaspectratio]{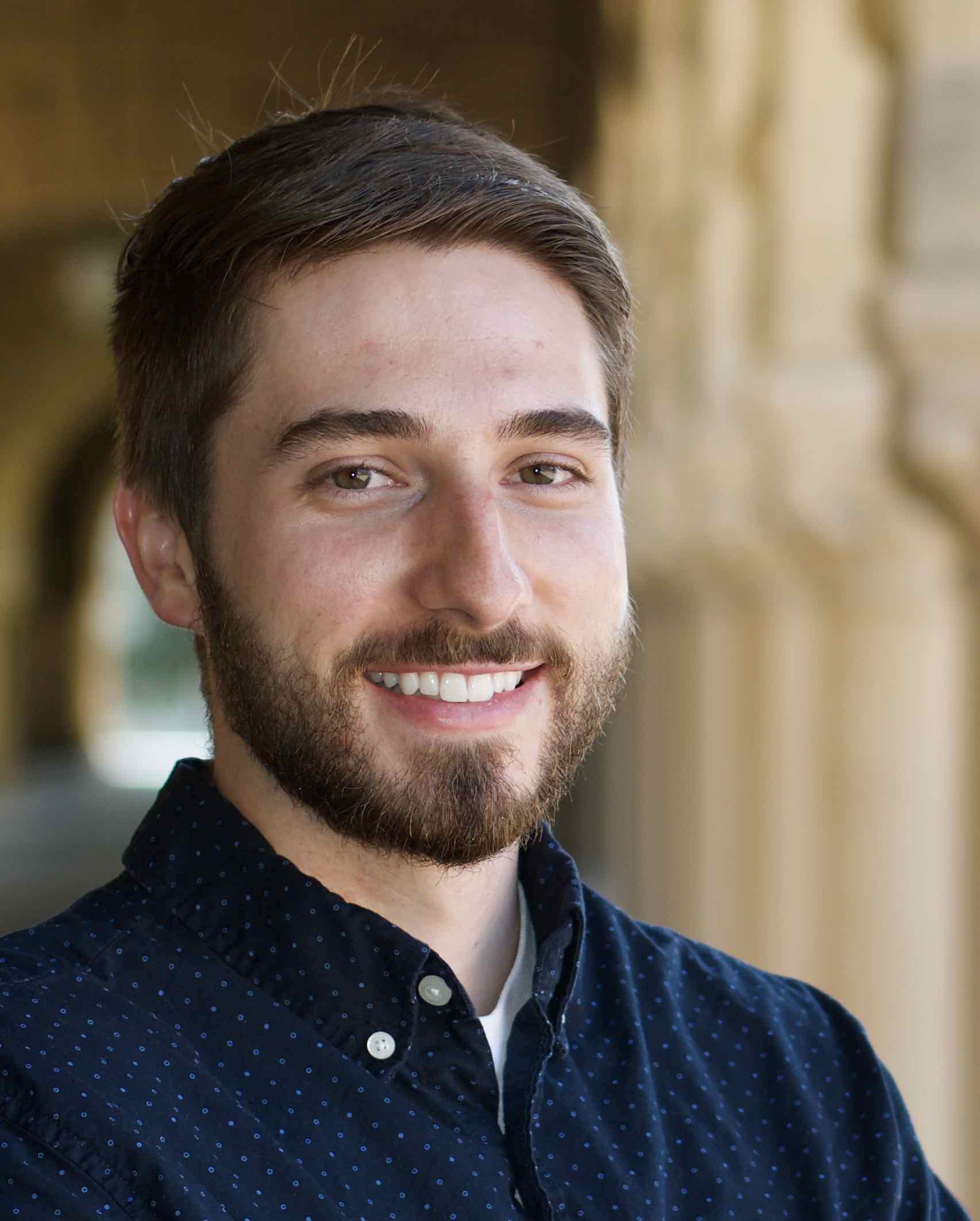}}]{Nathan Stacey}{\space} received a B.S. degree in mechanical engineering from Utah State University, Logan, UT in 2016. In 2018, he received an M.S. degree in aeronautics \& astronautics from Stanford University, Stanford, CA where he is currently pursuing a Ph.D. degree also in aeronautics \& astronautics. 

Nathan has completed internships with the Northrop Grumman Corporation and Space Dynamics Laboratory. Presently, he is a NASA Pathways intern at Goddard Space Flight Center, Greenbelt, MD and does research in the Stanford Space Rendezvous Laboratory. His research focuses on developing advanced estimation techniques for autonomous orbit determination with application to small celestial body missions. 

Mr. Stacey is the 2016 Utah State University Scholar of the Year, a National Science Foundation Graduate Research Fellow, a Stanford Enhancing Diversity in Graduate Education Doctoral Fellow, and an Achievement Rewards for College Scientists Scholar.
\end{IEEEbiography}

\begin{IEEEbiography}[{\includegraphics[width=1in,height=1.25in,clip,keepaspectratio]{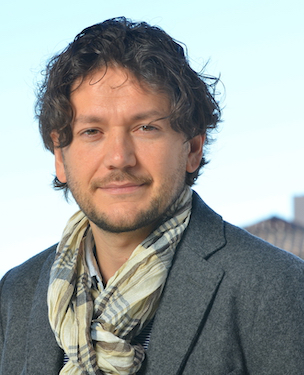}}]{Simone D'Amico} received his B.S. and M.S. degrees from Politecnico di Milano, Milan, Italy (2003) and his Ph.D. degree from Delft University of Technology, Delft, Netherlands (2010) in aerospace engineering. He is Associate Professor of Aeronautics and Astronautics (AA) at Stanford University and W. M. Keck Faculty Scholar in the School of Engineering. 

From 2003 to 2014, he was research scientist and team leader at DLR. There he made key contributions to the design, development, and operations of spacecraft formation-flying and rendezvous missions such as GRACE, TanDEM-X, PRISMA, BIROS, and PROBA-3. At Stanford, he is the founding Director of the Space Rendezvous Laboratory and Director of the AA Undergraduate Program. Dr. D'Amico's research aims at enabling future miniature distributed space systems for unprecedented science and exploration. His efforts lie at the intersection of advanced astrodynamics, GN\&C, and space system engineering to meet their tight requirements. As such, he is PI of the GN\&C system of several upcoming autonomous satellite swarms missions such as STARLING, SWARM-EX, and VISORS. He is an Associate Fellow of AIAA, Associate Editor of AIAA JGCD, and Chairman of the NASA Starshade Technology Working Group. 

Dr. D’Amico was the recipient of several awards, including Best Paper Awards at IEEE (2021), AIAA (2021), and AAS (2019) conferences, the Leonardo 500 Award by the Leonardo da Vinci Society/ISSNAF (2019), FAI/NAA’s Group Diploma of Honor (2018), DLR’s Sabbatical/Forschungssemester (2012) and Wissenschaft Preis (2006), and NASA’s Group Achievement Award for the GRACE mission (2004). More: https://damicos.people.stanford.edu/

\end{IEEEbiography}
\end{document}